\documentclass[aip, cha, amsmath, amssymb, reprint]{revtex4-1}

\usepackage{siunitx}
\usepackage{hyperref}
\usepackage{graphicx}
\usepackage{dcolumn}
\usepackage{bm}
\usepackage[utf8]{inputenc}
\usepackage[T1]{fontenc}
\usepackage{mathptmx}
\usepackage{etoolbox}
\usepackage{comment}
\usepackage{braket}
\usepackage[draft]{fixme} 
\usepackage{booktabs}   
\usepackage[shortlabels]{enumitem}
\setlist[enumerate]{noitemsep}

\fxsetup{layout=inline}
\FXRegisterAuthor{ceh}{cEH}{Carlos}
\FXRegisterAuthor{ib}{iB}{Ian}
\FXRegisterAuthor{pf}{pF}{Petros}
\FXRegisterAuthor{ag}{aG}{Adam}

\renewcommand{\selectlanguage}[1]{}

\makeatletter
\def\@email#1#2{%
 \endgroup
 \patchcmd{\titleblock@produce}
  {\frontmatter@RRAPformat}
  {\frontmatter@RRAPformat{\produce@RRAP{*#1\href{mailto:#2}{#2}}}\frontmatter@RRAPformat}
  {}{}
}%
\makeatother

\begin{document}
\preprint{AIP/123-QED}

\title[Single-photon emitters and spin-photon interfaces in silicon]{Single-photon emitters and spin-photon interfaces in silicon}

\author{Kilian Sandholzer}
\affiliation{Technical University of Munich, TUM School of Natural Sciences, Department of Physics and Munich Center for Quantum Science and Technology (MCQST), James-Franck-Stra{\ss}e 1, 85748 Garching, Germany}
\affiliation{TUM Center for Quantum Engineering (ZQE), Am Coulombwall 3A, 85748 Garching, Germany}

\author{Ian Berkman}
\affiliation{Research Laboratory of Electronics, Massachusetts Institute of Technology, Cambridge 02139, USA}

\author{Peter De\'ak}
\affiliation{HUN-REN Wigner Research Centre for Physics, P.O.\ Box 49, H-1525 Budapest, Hungary}

\author{Carlos Errando-Herranz}
\affiliation{QuTech and Kavli Institute of Nanoscience, Delft University of Technology, Delft 2628 CJ, Netherlands}
\affiliation{Department of Quantum and Computer Engineering, Delft University of Technology, Delft 2628 CJ, Netherlands}

\author{Petros-Panagis Filippatos}
\affiliation{HUN-REN Wigner Research Centre for Physics, P.O.\ Box 49, H-1525 Budapest, Hungary}

\author{Adam Gali}
\affiliation{HUN-REN Wigner Research Centre for Physics, P.O.\ Box 49, H-1525 Budapest, Hungary}
\affiliation{Department of Atomic Physics, Institute of Physics, Budapest University of Technology and Economics,  M\H{u}egyetem rakpart 3., H-1111 Budapest, Hungary}
\affiliation{MTA–WFK Lend\"{u}let "Momentum" Semiconductor Nanostructures Research Group, P.O.\ Box 49, H-1525 Budapest, Hungary}

\author{Andreas Gritsch}
\affiliation{Technical University of Munich, TUM School of Natural Sciences, Department of Physics and Munich Center for Quantum Science and Technology (MCQST), James-Franck-Stra{\ss}e 1, 85748 Garching, Germany}
\affiliation{TUM Center for Quantum Engineering (ZQE), Am Coulombwall 3A, 85748 Garching, Germany}

\author{Andreas Reiserer}
\email[]{andreas.reiserer@tum.de}
\affiliation{Technical University of Munich, TUM School of Natural Sciences, Department of Physics and Munich Center for Quantum Science and Technology (MCQST), James-Franck-Stra{\ss}e 1, 85748 Garching, Germany}
\affiliation{TUM Center for Quantum Engineering (ZQE), Am Coulombwall 3A, 85748 Garching, Germany}

\date{\today}

\begin{abstract}
Single photons enable the distribution of quantum information over large distances and thus play a major role in quantum technologies such as communication and computing. Solid-state emitters are practical and efficient sources of single photons that can be manufactured in large numbers. When combined with a spin, the resulting spin-photon interfaces can store quantum states for extended periods and serve as the basis for quantum networks and repeaters. Among the many host materials explored over the past few decades, silicon stands out for its advanced nanofabrication, the maturity of its integrated photonics and microelectronics, and its high isotopic purity, which leads to exceptionally long spin coherence. These properties position silicon single-photon emitters and spin-photon interfaces among the most promising hardware platforms for implementing quantum networks and distributed quantum information processors. This review summarizes the current state of the art and open challenges towards coherent single-photon sources and scalable spin-photon interfaces based on color centers and erbium dopants in nanophotonic silicon structures.
\end{abstract}

\maketitle

\tableofcontents

\section{Introduction} \label{sec:introduction}

The quantum revolution began more than a century ago, when Planck speculated that energy exchange between light and matter occurs in discrete quanta, and Einstein concluded that a light beam consists of a stream of photons. Decades later, single-photon and photon-pair sources were used to demonstrate the wave-particle duality and the generation of non-classical correlations and entanglement~\cite{pan_multiphoton_2012}. Combining such sources in multi-photon interference experiments enabled the implementation of photonic quantum gates~\cite{obrien_demonstration_2003} and quantum teleportation, which serves as a fundamental ingredient for many quantum applications~\cite{pirandola_advances_2015}. 

Many years after these pioneering experiments, optical and near-infrared photons remain a crucial component of quantum technology. Such photons enable the transmission of quantum information at the speed of light via free-space and fiber-optical channels, with negligible decoherence even at room temperature. This is the basis for their application in quantum communication~\cite{gisin_quantum_2002} and quantum networking~\cite{duan_colloquium_2010, reiserer_cavity-based_2015, wehner_quantum_2018}. At the same time, the high fidelity and ease of photon manipulation with linear optical components is a powerful resource for photonic quantum computation~\cite{rudolph_why_2017, slussarenko_photonic_2019, alexander_manufacturable_2025, maring_versatile_2024}.

However, the rapid propagation of light also has a downside: it makes it difficult to store quantum states for extended periods. Even in the lowest-loss optical fibers, quantum information encoded in photons will be lost on a timescale of tens to hundreds of microseconds. This can be overcome by photonic quantum memories~\cite{afzelius_quantum_2015} or by spin-photon interfaces. The latter enables hybrid approaches to distributed quantum information processing, in which information is stored and processed in stationary quantum nodes connected by photonic channels~\cite{yan_silicon_2021}. The high connectivity between quantum nodes in this approach offers significant advantages for quantum error correction, potentially enabling scalable quantum information processing hardware~\cite{simmons_scalable_2024}.

To connect remote nodes, the quantum properties of the photonic resource states can be encoded in either continuous variables~\cite{braunstein_quantum_2005, weedbrook_gaussian_2012}, such as the field quadratures, or in discrete variables, given by the number of energy quanta. In this review, we will focus on the latter approach, specifically on the implementation of single-photon sources and spin-photon interfaces for quantum applications. In this context, pioneering experiments were based on atoms trapped in a vacuum, which offer exceptional control over optical and spin properties. High photon generation efficiencies have been achieved by resonator integration~\cite{reiserer_cavity-based_2015}. However, operating these sources places high demands on the experiments and requires complex setups that are challenging to integrate and deploy. Therefore, solid-state alternatives have been explored over the last decades as a promising alternative for up-scaling~\cite{aharonovich_solid-state_2016, senellart_high-performance_2017, fox_solid-state_2025, holewa_solid-state_2025}. The following sections will summarize the requirements that the corresponding systems must fulfill.

\subsection{An ideal single-photon emitter} \label{subsec:ideal_photonemitter}
For the mentioned quantum applications, an ideal photon emitter should facilitate the on-demand creation of a single excitation in a perfectly defined optical field mode, as sketched in Fig.~\ref{fig:SPS_and_spin-photon_interfaces}. In more detail, this means it should possess the following set of properties:

\begin{figure}[tb]
\includegraphics[width=1.0\columnwidth]{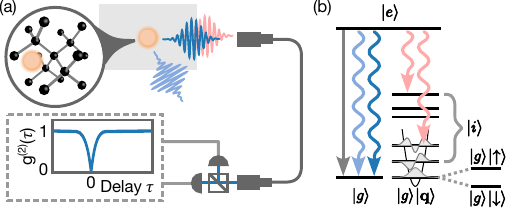}
\caption{     \label{fig:SPS_and_spin-photon_interfaces}
\textbf{Single-photon source and spin-photon interface.} (a) A single-photon source is realized by embedding a single emitter (orange) in a solid-state host (grey box and zoom-in to the silicon lattice). After excitation, the system emits light (curly arrows) via optical decay to the ground state. In an ideal system, all emissions occur into a well-defined mode of the electromagnetic field, which is then coupled to a detection setup (bottom), e.g., via optical fibers (grey). The single-photon nature of the field is verified by autocorrelation measurements (inset) in a Hanbury-Brown and Twiss setup. However, real emitters in bulk crystals will emit at multiple frequencies (red and blue) and modes (light blue). This limitation can be overcome by embedding the emitters into nanostructures. (b) In a spin-photon interface, the emitters have several ground-state spin levels, e.g., $\ket{\uparrow}$ and $\ket{\downarrow}$, to facilitate long-term storage of quantum information. Ideally, following excitation, the emitters will only decay radiatively to a single ground state in a mode that is efficiently collected (dark blue wavy arrow). In real devices, some photons may be lost (light blue), and the excited state may decay non-radiatively (grey arrow). Furthermore, the emitter may optically decay to other energy levels than its ground state $\ket{g}$, including other orbital states $\ket{i}$ or states with excited phonon modes $\ket{q}$. This leads to the emission of photons at different frequencies (red curly arrows). }
\end{figure}

\begin{enumerate}
\item It should exhibit a high radiative efficiency, concentrated on a single transition, which for solid-state sources means a negligible probability of non-radiative or phonon-assisted decays.
\item The source should emit only single photons; that is, the probability of simultaneously emitting multiple photons should be zero. 
\item All photons should be emitted in well-defined spatio-temporal and polarization modes, and the wavepacket duration should balance between fast and controllable operation timescales. 
\item The emission should be coherent and occur at a stable, constant frequency. This is a key requirement for implementing high-fidelity operations between independent sources using linear optical elements. Tunability of such frequency is highly desired for multi-source operation and spectral multiplexing.
\item The emitter should not exhibit blinking or bleaching, which would render it unavailable for photon generation during certain time intervals or after a certain usage time. 
\item The emission should be at a frequency that enables transmission over the targeted distances; for long-distance applications using optical fibers, this translates to operation in the minimal-loss bands of the optical telecommunications window~\cite{holewa_solid-state_2025}, while for local and datacenter-scale networks, a broader frequency range is possible, see Fig.~\ref{fig:overview}b. 
\item The fabrication of the emitters should be scalable, ideally using established wafer-scale manufacturing processes with high yield.
\item The devices should operate in a conveniently accessible temperature range. As coherent operation under ambient conditions has not been demonstrated with any solid-state emitter so far, this condition may need to include temperatures attainable with widely available commercial cryogenic hardware.
\item If the single-photon source is to be integrated into a spin-photon interface, as described in Sec.~\ref{subsec:ideal_spin-photon_interface}, the optical transition must couple efficiently to a spin degree of freedom with long coherence times.
\end{enumerate}

Due to the numerous requirements, implementing an ideal single-photon source is a complex and challenging task. Therefore, many different approaches are being explored. This includes probabilistic sources, such as those based on spontaneous parametric down-conversion~\cite{wang_integrated_2021} and spontaneous four-wave mixing~\cite{wang_progress_2024}, which are relatively easy to operate and to produce. However, this does not provide access to long-lived memories, which would then require heterogeneous integration of cold or warm ensembles of atoms or rare-earth doped solid-state crystals. Furthermore, the efficiency of such photon sources is fundamentally limited by multi-photon emission~\cite{barbieri_parametric_2009, christ_limits_2012}, and extending them towards the production of multi-photon entangled states is conceivable~\cite{rudolph_why_2017}, but challenging. In contrast, deterministic photon sources based on single emitters enable the simultaneous achievement of high efficiency and high single-photon purity.

\subsection{An ideal spin-photon interface} \label{subsec:ideal_spin-photon_interface}
While such emitter-based deterministic single-photon sources facilitate the distribution of quantum states, spin-photon interfaces add quantum memory and deterministic processing capabilities, thereby offering additional possibilities. As examples, spin-photon interfaces enable the efficient generation of multi-photon entangled states~\cite{thomas_efficient_2022, cogan_deterministic_2023}, and may pave the way towards quantum repeaters~\cite{briegel_quantum_1998} and networks~\cite{wehner_quantum_2018}, as well as distributed quantum information processors~\cite{simmons_scalable_2024}. 

For these purposes, a spin-photon interface should allow ideal single-photon generation according to the criteria in Sec.~\ref{subsec:ideal_photonemitter} and fulfill the following, additional properties: 
\begin{enumerate}[noitemsep]
\item An ideal spin-photon interface should enable the optical initialization and readout of a spin qubit, as well as the efficient generation of spin-photon entanglement, either by photon emission or by spin-photon quantum gates~\cite{reiserer_colloquium_2022}.
\item The device should facilitate single-shot readout of the spin state. To this end, it may exhibit an optical transition with high cyclicity, meaning the spin state is not altered upon repeated optical excitation and decay.
\item The spin-photon interface should give access to a larger register of several spins~\cite{dutt_quantum_2007, robledo_highfidelity_2011} with controlled interactions, allowing the implementation of quantum gates with high fidelity. This register should exhibit long coherence times, even when some of the spins are excited optically.
\item Between the spins of these registers, fast, high-fidelity control should be facilitated either all-optically or via microwave or radio-frequency pulses, in a frequency regime easily accessible to electronic devices.
\end{enumerate}

While atoms trapped in a vacuum enable the implementation of such spin-photon interfaces~\cite{reiserer_cavity-based_2015}, solid-state systems may enable reduced experimental overhead and higher qubit densities~\cite{vandersypen_interfacing_2017}, while avoiding limited qubit lifetimes from background-gas scattering. In addition, the coherence of nuclear spins in solids can reach the timescale of hours~\cite{saeedi_room-temperature_2013, zhong_optically_2015, wang_nuclear_2025}, exceeding that of any other platform, and further improvements will likely be possible.

The dominant limitations of the coherence time are spin-lattice relaxation and the interaction of spins in solids with one another. Both can be minimized by an appropriate choice of materials and operating conditions in terms of temperature and magnetic field~\cite{wolfowicz_quantum_2021}. In addition, spin-spin interactions can be mitigated by tailored dynamical decoupling protocols~\cite{suter_colloquium_2016, zhou_robust_2023}, which enable enhancing the coherence of individual spin qubits by many orders of magnitude~\cite{abobeih_onesecond_2018}. Finally, if sufficiently high control fidelities can be achieved in the future, tailored quantum error correction protocols~\cite{breuckmann_quantum_2021} can be used to eliminate the remaining limitations of qubit coherence in spin-photon interfaces. 

\begin{figure*}[tb]
\includegraphics[width=2.0\columnwidth]{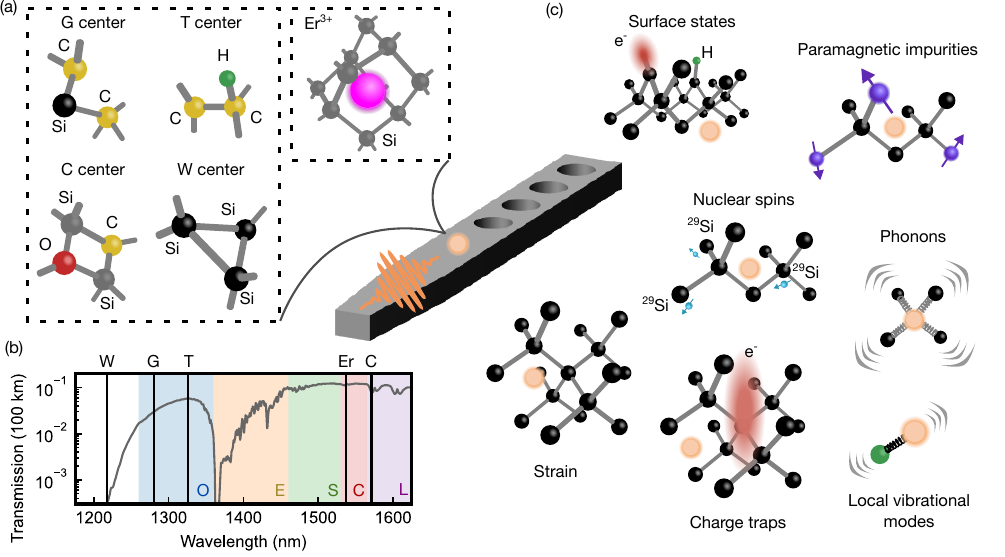}
\caption{     \label{fig:overview}
\textbf{Single-photon emitters embedded in nanophotonic silicon devices.} (a) The most prominent color centers are named by letters (G, T, C and W centers). They are formed by specific arrangements of lattice impurities such as Carbon (C), Oxygen (O), and Hydrogen (H), as well as interstitial Silicon (Si) atoms. Their energy levels have a significant contribution from the orbitals of the surrounding lattice sites. In contrast, Erbium dopants, typically in their trivalent charge state ($\text{Er}^{3+}$), exhibit optical transitions between energy levels in their partially filled inner 4f-shell, making them less sensitive to lattice perturbations. (b) The emission wavelength of the emitters shown in (a) falls within the major telecommunication bands (O-, E-, S-, C-, and L-band; colors). Here, the transmission of light through low-loss optical fibers can still approach $10\%$ after \SI{100}{\kilo\meter} (grey data from~\cite{petrovich_broadband_2025}). (c) The optical stability of the energy levels, and thus the coherence of the emitted photons, is limited by various sources of decoherence. Charge traps at the surface and in bulk can lead to a fluctuating charge environment. In addition, fluctuating magnetic fields can arise from paramagnetic impurities and the nuclear spin bath. Finally, coupled local and extended phonon modes, as well as external vibrations, may also reduce the coherence.}
\end{figure*}

This review describes how the mentioned challenges can be overcome using color centers and erbium dopants in nanophotonic silicon devices, as shown in Fig.~\ref{fig:overview}a. To this end, we will first explain the basic properties of solid-state photon emitters in Sec.~\ref{sec:emitters_and_SPIs}. This will be followed by a detailed discussion of the properties of the -- as of today -- most advanced silicon emitters: Erbium dopants and the T, W, C, and G centers in Sec.~\ref{sec:silicon_emitters}. While ensembles of such emitters can be used for spectroscopic investigations and applications such as quantum memories and microwave-to-optical transducers~\cite{higginbottom_memory_2023, khalifa_robust_2025, rinner_quantum_2026}, this review will focus on single emitters and their integration into nanophotonic devices to enable efficient, coherent spin-photon interfaces in Sec.~\ref{sec:Emitters_in_nanostructures}. Based on this, we will summarize recent achievements and sketch the route towards practical scalability in Sec.~\ref{sec:practical_scalability}, with a focus on quantum technological applications in sensing, communication, networking, as well as photonic and distributed quantum information processing. Finally, we will outline future directions in this emerging research field.

\section{Solid-state single-photon emitters and spin-photon interfaces}   \label{sec:emitters_and_SPIs}
The implementation of single-photon sources and spin-photon interfaces that fulfill all of the challenging requirements outlined in Sec.~\ref{subsec:ideal_photonemitter} and \ref{subsec:ideal_spin-photon_interface} is still an open challenge. The following sections will summarize the key limitations encountered in this context with solid-state systems.

\subsection{Spectral stability of solid-state emitters}
To enable coherent operation, the emitters must be isolated from interactions with their environment. For an emitter in a solid-state host matrix, this is a fundamental challenge, as nearby atoms or impurities can easily perturb the emitter via electric, magnetic, or strain fields. This leads to fluctuations of the optical and spin transition frequencies, which are typically distinguished by their timescales. In a static-disorder environment, the energy levels of different emitters exhibit random yet static shifts, an effect known as inhomogeneous broadening. Additionally, the  transition frequencies may be affected by fields that change over timescales that are long or short compared to the lifetime; the former effect is known as spectral diffusion, while the latter is referred to as homogeneous broadening or dephasing.

The three mentioned effects have different impacts on the devices' performance as single-photon sources or spin-photon interfaces. A moderate static disorder is typically not a major concern: On the spin transitions, it can be overcome by dynamical decoupling~\cite{suter_colloquium_2016}, provided single-qubit rotations with a high fidelity can be implemented. Inhomogeneous broadening of the optical transition frequency can also be counteracted, e.g., by frequency shifting elements~\cite{levonian_optical_2022} or by time-gated photon detection~\cite{vittorini_entanglement_2014}. As it enables spectral multiplexing~\cite{chen_parallel_2020} of up to hundreds of emitters~\cite{ulanowski_spectral_2024}, it can even be employed as a powerful resource in quantum information processing and quantum networking~\cite{ruskuc_multiplexed_2025}. In contrast, spectral diffusion and homogeneous broadening beyond the lifetime-limited linewidth are always detrimental for quantum applications. While the former can be overcome by tailored decoupling protocols, both on the spin~\cite{suter_colloquium_2016} and on the optical~\cite{ruskuc_multiplexed_2025, uysal_rephasing_2024} transitions, the latter will inevitably reduce the achievable fidelity and/or the rate of quantum operations. Therefore, it is paramount to understand and control the spectral diffusion and homogeneous broadening of quantum emitters. 

The dominant sources of noise encountered by photon emitters in nanostructures are summarized Fig.~\ref{fig:overview}c. Those include paramagnetic impurities, charge traps, surface states, nuclear spins, strain fields, phonons, and local vibrational modes. While some of these noise sources are encountered even in the purest bulk crystals studied to date, others are relevant only in nanostructures, as their influence scales with distance $r$. As an example, the electric field $E$ generated by a fluctuating charge environment (e.g., caused by trap states on the surface or in the bulk) will scale as $E_\text{charge}\propto r^{-2}$. In comparison, the electric or magnetic fields caused by the dipoles of paramagnetic impurities, nuclear spins, or two-level systems will drop much faster, as $E_\text{dipole}, B_\text{dipole}\propto r^{-3}$. In addition to these local noise sources, fluctuations in static offset fields and vibrations that distort the crystalline environment, leading to time-varying strain, can induce spectral instabilities and thus require control. In the following, we will discuss the relevance of these perturbations on the different transitions of the emitters.

We start with the spin levels, which are typically not sensitive to electric but only to magnetic fields. Thus, their stability is mostly limited by dipolar magnetic spin-spin interactions, and the best coherence can be achieved in a "magnetic vacuum", i.e., a crystal that is free of paramagnetic impurities and nuclear spins to the largest degree possible. In addition, qubits with a low sensitivity to magnetic fields will exhibit better coherence. This favors nuclear spins compared to their electronic counterparts, as the magnetic moment of a spin scales with its mass. Thus, nuclear-spin transitions exhibit a much weaker magnetic-field dependence, typically by four orders of magnitude. Even lower values can be achieved in hyperfine states, i.e., in coupled electron-nuclear states that can exhibit a zero-first-order Zeeman (ZEFOZ) shift under certain external fields~\cite{zhong_optically_2015, ortu_simultaneous_2018}, paving the way to hour-long coherence even in host materials with a significant density of fluctuating nuclear spins~\cite{zhong_optically_2015}.

Fluctuations in the spin-level energies can also affect the optical emission. However, typically this is not the dominant limitation: unlike bare spin transitions within the same orbital, optical transitions involve different orbitals. Thus, they are easily perturbed by electric fields, which is known as the Stark effect. While external electric fields can be shielded efficiently, charge noise within the solid-state environment of the emitters remains a major concern. Thus, good optical coherence is achieved only in crystals with a low concentration of fluctuating charges, which typically requires high chemical and structural crystalline quality. However, one can try to stabilize the charge environment by applying large static electric fields~\cite{lodahl_interfacing_2015, anderson_electrical_2019}. 

In addition to minimizing the fluctuations, one can reduce the sensitivity of the transitions to electric fields by using emitters that do not exhibit a permanent electric dipole moment, and thus no first-order Stark shift~\cite{macfarlane_optical_2007,kaplyanskii_linear_2002}. This can be achieved in emitters with non-polar point symmetry; in silicon, this includes, e.g., the $D_{3d}$ symmetry of hexagonal and bond-centered interstitial sites, and the $T_d$ symmetry of tetrahedral-interstitial and substitutional sites. Furthermore, in rare-earth emitters such as erbium, owing to the shielding of the inner 4f electrons by the filled outer 5s and 5p shells, the magnitude of the Stark coefficient is expected to be an order of magnitude weaker than that found in color centers with the same symmetry~\cite{macfarlane_optical_2007}.

Implementing a photon source with good optical coherence thus requires, first, emitters with minimal sensitivity to perturbations, and second, engineering the system to minimize these perturbations. In Sec.~\ref{sec:silicon_emitters}, we will describe how these two goals can be achieved with single-photon emitters in silicon.

\subsection{Efficient and coherent photon generation in a single mode of the electromagnetic field}
Besides its spectral stability, a key figure of merit for single-photon sources or spin-photon interfaces is efficiency, defined as the optical power that couples into a given target optical mode relative to the total emitted power. In most practical cases, the target mode is defined by a single-mode optical fiber or a nanophotonic waveguide. In bulk crystals, however, the optical modes to which the generated photons are coupled are determined solely by the emitters, hindering efficient collection to a single-mode fiber. Furthermore, light extraction can be hampered by total internal reflection, particularly in solids with a high refractive index such as silicon, with $n\simeq 3.45$. Numerous approaches exist to still enable efficient photon collection~\cite{aharonovich_solid-state_2016, fox_solid-state_2025, reiserer_colloquium_2022}, which will be described in detail in Sec.~\ref{sec:Emitters_in_nanostructures}; most of them are based on integrating the emitter into nanostructures.

Integrating photon emitters into nanostructures can not only improve the light extraction into a certain mode; it also allows modifying the optical transition rates by tailoring the photonic local density of states (LDOS)~\cite{lodahl_interfacing_2015, reiserer_colloquium_2022} $\rho (\omega)$. In this context, one needs to consider that quantum applications require efficient single-frequency photon generation. However, the emitters may not only decay at a rate $R_\text{g}$ to a single ground state $\ket{g}$ while generating photons at a specific transition frequency $\nu_\text{g}$ into the desired light mode; they may also decay into different spatial modes, or to different electronic energy levels, at rates $R_i$ and frequencies $\nu_i$. In addition, the emitters may exhibit phonon-sideband (PSB) decays at $R_\text{PSB}$, emitting light at a smaller frequency while simultaneously exciting a phonon mode. Finally, there may be non-radiative decay pathways at a rate $R_\text{NR}$ that reduce the overall photon emission efficiency.

The sum of these effects determines the rate $R$ at which the emitters decay from their optically excited state $\ket{e}$:

\begin{equation}  \label{eq:DecayRate}
R = R_\text{g} + \sum_i R_i + R_\text{PSB} + R_\text{NR}
\end{equation}

Here, only the first term accounts for the radiative rate of the desired transition to the ground state. Thus, to implement an efficient single-photon source or a spin-photon interface, one must ensure that the first term is much larger than all the others. In bulk crystals, this means that the emitters must exhibit negligible non-radiative decay, i.e., high quantum efficiency (sometimes also called quantum yield), defined as 

\begin{equation}  \label{eq:Quantum_efficiency}
\eta_\text{QE}=\frac{R-R_\text{NR}}{R}. 
\end{equation}  

In addition, the emitters should exhibit a low $R_{PSB}$, such that photons are only emitted at the zero-phonon line (ZPL) $R_\text{ZPL}= R_\text{g} + \sum_i R_i$. This condition is quantified by the Debye-Waller-Factor (DWF):

\begin{equation}  \label{eq:DWF}
\text{DWF}=\frac{R_\text{g} + \sum_i R_i}{1-R_\text{NR}}=\frac{R_\text{ZPL}}{R_\text{ZPL}+R_\text{PSB}}. 
\end{equation} 

Finally, the emitters should exhibit a high branching ratio, which measures the relative strength of the radiative rate for the desired transition compared to all others in the ZPL:
\begin{equation}  \label{eq:Branching_ratio}
\eta_\text{BR}= \frac{R_g}{\sum_i R_i}. 
\end{equation}  

Clearly, a high branching ratio is required to achieve high contrast in photon interference measurements when the different ZPL transitions exhibit different polarizations, frequencies, and/or rates. Furthermore, in spin-photon interfaces, owing to the limitations of photon detectors and the finite outcoupling efficiency of typical devices, a high branching ratio is advantageous for single-shot spin readout, as it allows multiple excitations without altering the spin state. In this context, it may also be sufficient to have a high cyclicity, which --- independent of the decay path --- quantifies the number of optical excitations before the initial spin state is lost.

In summary, the finite values of $\eta_\text{BR}$, DWF, and $\eta_\text{QE}$ in bulk crystals hamper the efficiency of photon generation into a single mode. However, tailoring the optical density of states in nanostructures allows one to overcome this limitation, as explained in more detail in Sec.~\ref{sec:Emitters_in_nanostructures}. In effect, the efficiency can be boosted by suppressing the emission rate on the undesired transitions~\cite{yablonovitch_inhibited_1987} and/or by increasing that on the desired transition~\cite{purcell_spontaneous_1946}.

So far, most experiments have focused on the latter approach, i.e., increasing $R_g$ such that $R_\text{g} \gg \sum_i R_i + R_\text{PSB} + R_\text{NR}$. This approach leads to a reduction of the optical lifetime $T_1=1/R$ compared to the bulk value $1/R_0$, which is quantified by the Purcell enhancement factor~\cite{reiserer_colloquium_2022} $F_\text{P}$:

\begin{equation}  \label{eq:PurcellFactor_definition_via_rates}
F_\text{P}= R/R_0
\end{equation}

This definition generalizes the Purcell factor, originally defined for resonators~\cite{purcell_spontaneous_1946}, to arbitrary photonic structures. $F_\text{P}$, as defined above, can be directly determined by measurements; however, note that parts of the literature use different definitions.

In case $F_\text{P}$ is much larger than one, $F_\text{P}\cdot R_0 \gg R_i, R_\text{PSB}, R_\text{NR}$, such that the detrimental effects of finite branching, phonon-sideband emission and non-radiative decay can be overcome. Thus, nanopatterning can significantly enhance the efficiency of single-photon sources and spin-photon interfaces while simultaneously accelerating their emission. 

Remarkably, the resulting lifetime reduction can also relax the requirements on the emitters' optical coherence time $T_2$. This is a key concern for single-photon sources and spin-photon interfaces: Only if $T_2$ is limited by the emitter lifetime $T_1$, i.e., $T_2=2T_1$, the emitted photons achieve optimal performance for quantum information processing. However, in solids, additional dephasing processes can strongly reduce $T_2$. The basic idea is thus to shorten $T_1$ by the Purcell effect such that the Fourier limit can be attained. In other words, the radiative linewidth (which is determined by the inverse of the emitter lifetime) is broadened such that it becomes larger than the homogeneous linewidth. In the following, this will be explained in more detail.

In the radiative decay process, the phase of the emitted light is determined by the emitter polarization; thus, the decay rate of the emitter polarization $\gamma_\perp$ determines the optical coherence time. $\gamma_\perp$ has contributions from the desired radiative transition (in bulk at a rate $\gamma_g=R_\text{g}/2$), the other decay paths $\gamma_1=(R-R_\text{g})/2$, and dephasing $\gamma_\text{d}$ caused by energy level fluctuations, i.e., $\gamma_\perp=\gamma_\text{g}+\gamma_1+\gamma_\text{d}$. In bulk solids, the latter is typically the largest term, such that $\gamma_\perp \simeq \gamma_\text{d}$. In this case, the emitted light is incoherent. 

In contrast, integration into nanophotonic devices can lead to a fast radiative decay into the desired mode at a rate $\gamma_c$. To achieve coherent emission, one thus needs to fulfill $\gamma_\text{c} = F_\text{P} \cdot \gamma_\text{g} \gg \gamma_\text{d}$. This can be written as:

\begin{equation}    \label{eq:Cooperativity_vs_Purcell}
C \simeq F_\text{P}\frac{\gamma_\text{g}}{\gamma_\text{d}} \gg 1
\end{equation}

The quantity $C$ is the cooperativity known from cavity quantum electrodynamics. It can be calculated from the emitter-field coupling rate $g$ and the field decay rate at the emitter position $\kappa$, which gives~\cite{reiserer_cavity-based_2015}:

\begin{equation} \label{eq:cooperativity}
    C = \frac{g^2}{2 \kappa \gamma_\perp}.
\end{equation}

Note that other publications may use the energy rather than the field and polarization decay rates ($K=2\kappa$ and $\Gamma = 2\gamma$), and/or the vacuum Rabi frequency $\Omega=2g$ instead of the coupling rate; this can give corresponding factors of 2 in the above equations~\cite{janitz_cavity_2020}. In Sec.~\ref{sec:Emitters_in_nanostructures}, it will be explained how the criterion $C\gg 1$ can be achieved with photon emitters in nanophotonic silicon devices.

\section{Single-photon emitters in silicon}     \label{sec:silicon_emitters}

The previous sections have summarized the challenges in developing ideal solid-state single-photon sources and spin-photon interfaces. Owing to the plethora of requirements, this remains an active area of research. In this review, we will focus on the use of silicon as a host for single-photon emitters towards this goal, which appears particularly promising for upscaling due to the unrivaled maturity of its nanofabrication. While these capabilities were originally developed for microelectronic devices, it has been realized that silicon's high purity and high refractive index also make it an excellent material for on-chip light confinement in low-loss photonic devices. Meanwhile, silicon photonic foundries offer device manufacturing as a commercial service, greatly easing access to such devices for both companies and research groups~\cite{shekhar_roadmapping_2024, rahim_open-access_2018}.

In addition to its upscaling potential, silicon can be grown almost free of paramagnetic defects. Purification in the nuclear-spin-free isotope $^{28}\text{Si}$ can further reduce magnetic noise density. This makes silicon an exceptional host for spin qubits, with nuclear spin qubits reaching coherence times on the timescale of hours~\cite{saeedi_room-temperature_2013}. Thus, implementing a spin-photon interface in silicon offers unique possibilities for quantum networking~\cite{wehner_quantum_2018, reiserer_colloquium_2022} and distributed quantum computing~\cite{yan_silicon_2021, simmons_scalable_2024}.

Despite the mentioned advantages of silicon, the implementation of single-photon sources and spin-photon interfaces into this mature photonics platform has only recently emerged, as both research groups and companies have increasingly focused on the scalability of quantum hardware. The reason is found in the band structure of silicon. The indirect nature of its bandgap has so far hindered the realization of quantum dots with promising photon emission properties. In addition, the relatively small bandgap prevents the generation of light in the visible range, where efficient, inexpensive, and low-noise photon detectors are available. Even in the telecommunications range, two-photon absorption may hinder the application of high-power optical control fields. In addition, emitters with such a low energy-level separation may be prone to non-radiative decay via phonons or defect states.

Despite these challenges, several emitters in silicon have been identified that may fulfill the requirements of efficient photon sources and spin-photon interfaces summarized in Sec.\ref{sec:introduction}. They fall into two categories: dopant atoms, whose energy levels originate in their internal electronic structure, and color centers, whose orbitals have a larger extent, allowing the wavefunctions of their eigenstates to exhibit significant contributions from nearby atoms, including both silicon and other impurities. An overview of the most prominent emitters, as well as their structure, is shown in Fig.~\ref{fig:overview}a. Their detailed properties, similarities, and differences will be explained in Sec.~\ref{subsec:ErbiumDopants} and \ref{subsec:ColorCenters}. Their near-infrared emission falls between the band-edge of silicon (around $\SI{1.1}{\micro\meter}$) and the L-band, as summarized in Fig.~\ref{fig:overview}c. For the emitters presented in this review, the photon energy is well below the host material's bandgap, ensuring that the light is not directly absorbed by the host material. Thus, all these emitters may serve as the basis for efficient single-photon sources. In addition, the exploration of new color centers and optically active defects in silicon is an active field of research. Recently studied emitters include, for example, alkali-metal-vacancy complexes~\cite{udvarhelyi_design_2025}, selenium dopants~\cite{morse_photonic_2017, deabreu_characterization_2019}, carbon-nitrogen~\cite{nangoi_carbon-nitrogen_2026, udvarhelyi_first-principles_2026}, carbon-interstitial~\cite{deak_quantum_2024}, and the M center~\cite{filippatos_reexamination_2025,filippatos_first-principles_2026}. Computational methods for identifying and understanding these emitters and others will be briefly outlined in Sec.~\ref{subsec:TheoreticalMethods_NewEmitters}.

\subsection{Silicon as host material} \label{subsec:SiliconHost}

Implementing single-photon sources with good optical coherence requires minimizing perturbing electric, magnetic, and strain fields. To this end, one will aim at integrating the emitters into silicon crystals with high chemical purity and structural quality. The growth of such crystals has been optimized over decades in semiconductor manufacturing, and comprehensive textbooks provide detailed summaries~\cite{hull_properties_1999}. The most commonly used technique is Czochralski (Cz) growth, in which the raw material is heated in a crucible, and the crystal is grown by slowly pulling a seed rod from the melt. The method achieves very high purity, with the dominant contamination being oxygen at typical concentrations on the order of $10^{18}\si{\centi\meter^{-3}}$. Thus, the average distance between oxygen atoms in such crystals is already $\sim\SI{10}{\nano\meter}$. Other common impurities are hydrogen and carbon at slightly lower concentrations.

Even higher purity is achieved by the crucible-free float-zone (FZ) technique. Here, contaminant atom concentrations below $10^{15}\si{\centi\meter^{-3}}$ are routinely achieved, corresponding to an average distance between impurities exceeding $\SI{100}{\nano\meter}$. In such bulk crystals, one can minimize strain perturbations to the embedded emitters, enabling ensembles to emit within a narrow spectral window. The remaining inhomogeneous linewidth depends on the strain sensitivity of the emitters~\cite{stoneham_shapes_1969}, but is typically $\lesssim\SI{20}{\giga\hertz}$ for color centers~\cite{davies_optical_1989} and $\lesssim\SI{1}{\giga\hertz}$ for erbium dopants \cite{gritsch_narrow_2022}. This frequency range can be bridged by optical modulators, allowing photons emitted by different sources to be easily shifted to the same frequency. 

A remaining source of broadening is the isotopic content of silicon. Crystals with natural abundance consist of the isotopes $^{28}\text{Si}$ with \SI{92.2}{\percent} abundance, $^{29}\text{Si}$ with $4.7\,\%$, and $^{30}\text{Si}$ with $3.1\,\%$. Because of the different weights of these isotopes, their zero-point fluctuations have a different energy, which leads to a strain inhomogeneity in natural abundance crystals~\cite{cardona_isotope_2005}. This can be strongly reduced by the growth of isotopically purified crystals, typically in the most abundant $^{28}\text{Si}$. In crystals with extremely low defect density and isotopic impurities, a hundredfold reduction of the inhomogeneous linewidth down to tens of MHz has been observed with some color centers~\cite{chartrand_highly_2018}.

So far, such outstanding spectral properties have only been achieved in bulk crystals. The reason is that additional sources of decoherence arise in nanofabricated devices. In particular, these include dangling bonds, surface states, and contaminant atoms at the interface, which are in close proximity to the emitters in the nanostructure. Additionally, crystal damage can occur during the fabrication of such devices, particularly in implantation processes commonly used for emitter integration~\cite{macquarrie_generating_2021, weiss_erbium_2021}. Here, one faces a general difficulty: achieving a high probability of forming a single emitter in a device with a nanoscale volume requires an emitter concentration significantly higher than in a bulk crystal, unless one resorts to device fabrication around previously determined emitter positions. Introducing color centers or erbium dopants at such high concentrations, however, can significantly harm the crystalline integrity and thus the spectral properties of the emitters.

Still, to achieve the best results, one typically starts fabrication of nanoscale devices with pure crystals, often in thin-film form. Most experiments utilize silicon-on-insulator wafers for this purpose, which are commercially available in sufficient quality and purity. They are often made using the smart-cut technique or the bond-and-etchback method. This allows the fabrication of device layers as thin as $\SI{30}{\nano\meter}$, which can serve as a seed for the growth of ultrapure and even isotopically pure~\cite{liu_28silicon--insulator_2022} layers by chemical vapor deposition (CVD) or molecular beam epitaxy (MBE). A direct comparison of CVD-grown samples with Smart-Cut layers made from CZ silicon and bond-and-etchback wafers from FZ silicon gave similar results~\cite{gritsch_narrow_2022}, which means that currently all of these techniques seem equally suited for hosting single-photon emitters.

\subsection{Erbium dopants} \label{subsec:ErbiumDopants}

In the following section, we will describe the basic properties of erbium dopants that make them suited as single-photon emitters and spin-photon interfaces in silicon. Erbium is a rare-earth element that occupies its triply ionized state in most solid-state host materials. In this configuration, the inner 4f shells are enclosed by fully occupied 5s and 5p electronic shells. This results in effective shielding, making electrons in the 4f shell largely insensitive to electric-field noise caused by the surrounding crystal. Therefore, the spectra of the optical emission on transitions between 4f levels are largely independent of the host material~\cite{thiel_rare-earth-doped_2011}. The energy levels of the emitters are thus determined by the free-ion Hamiltonian and are only slightly split and shifted by the crystal field (CF). The latter is caused by the surrounding host crystal and differs between silicon and other materials. As phonons induce fast transitions between these CF levels, quantum devices will require operation at cryogenic temperatures, such that only the lowest levels in the CF manifolds of ground- and optically excited states are occupied. 

\subsubsection{Energy level scheme and spin properties}

\begin{figure}[tb] 
\includegraphics[width=1.0\columnwidth]{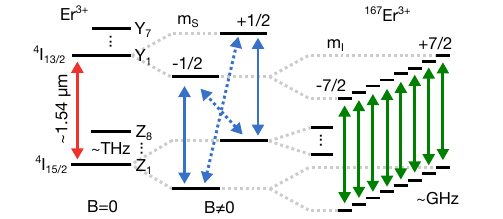}
\caption{     \label{fig:erbium_levelScheme}
\textbf{Level structure of erbium dopants in silicon}. Erbium exhibits eight crystal field levels ($Z_1,...Z_8$) in the ground state manifold ($^{4}I_{15/2}$) and seven ($Y_1,...Y_7$) in the excited state manifold ($^{4}I_{13/2}$). The levels are further split under an external magnetic field, forming an effective spin-1/2 system in the lowest crystal-field levels of each manifold. In addition to the electronic spin, the isotope $^{167}\mathrm{Er}$ exhibits a nuclear spin of 7/2, whose levels are split on the order of GHz via the hyperfine interaction. All levels can also be further split by the superhyperfine interaction with the nuclear spins surrounding the Er dopants (not shown). The optical transitions are indicated with arrows; one often distinguishes electron-spin-preserving (blue/green solid) and spin-flip (blue dashed) transitions. The nuclear-spin-flip transitions are not shown for simplicity.}
\end{figure}

The optical transition frequency of erbium is determined by the energy level separation between its $^4\text{I}_{15/2}$ ground state and its $^4\text{I}_{13/2}$ excited state. The wavelength of this transition typically ranges between $1530$ and $\SI{1540}{\nano\meter}$ and thus falls right into the minimal loss band of optical fibers, which is a key advantage for long-distance photon transmission. The frequency of the emitted light is around $\SI{195}{\tera\hertz}$ and thus exceeds the phonon frequencies more than tenfold, such that purely phononic relaxation plays no role, and one expects that the optical transitions will be dominated by radiative decay at cryogenic temperature. At higher temperatures, typically tens of Kelvin, cross-relaxation with impurities can be observed, reducing the lifetime of the optically excited state~\cite{priolo_excitation_1998, kenyon_erbium_2005, vinh_photonic_2009, sandholzer_luminescence_2025}. Already at lower temperatures, spin relaxation can dephase optical transitions. These effects can be used for accurate thermometry in nanoscale devices~\cite{sandholzer_luminescence_2025}. However, the resulting reduction in the coherence of the emitted light~\cite{gritsch_narrow_2022} so far hinders quantum applications at temperatures $\gtrsim\SI{8}{\kelvin}$.

Using Er:Si for the implementation of single-photon sources or spin-photon interfaces requires a detailed knowledge of its energy levels. The underlying structure, shown in Fig.~\ref{fig:erbium_levelScheme}, is very similar in all materials and sites owing to the shielding of the inner 4f levels. At cryogenic temperature, only the lowest CF level, $Z_1$, of the $^{4}I_{15/2}$ ground state manifold will be populated. The temperature required for coherent operation thus depends on the separation between $Z_1$ and $Z_2$, which is on the order of $\si{\tera\hertz}$ in low-symmetry sites~\cite{thiel_rare-earth-doped_2011}. 

The precise ordering and separation of the CF levels depend on the microscopic structure of the erbium sites. Heterogeneous strain in the crystal thus leads to an inhomogeneous broadening of the optical transitions~\cite{stoneham_shapes_1969}. This effect can be used to selectively address many individual erbium emitters in a nanoscale volume~\cite{chen_parallel_2020, gritsch_optical_2025} and to measure the strain in nanoscale devices~\cite{zhang_single_2019}. The inhomogeneous broadening can also be used to increase the bandwidth of optical quantum memories based on rare-earth dopants~\cite{afzelius_quantum_2015} through patterning of its absorption profile by persistent spectral hole burning, which can be achieved in Er:Si via its long-lived spin states.

As trivalent erbium is a Kramers dopant, meaning it has an odd number of electrons in the 4f shell and every CF level is two-fold degenerate at zero magnetic field~\cite{liu_spectroscopic_2005}. Upon the application of a static magnetic field $\vec{B}$, these energy levels split due to the Zeeman effect, as shown at the center of Fig.~\ref{fig:erbium_levelScheme}. At low fields, the splitting of the individual CF levels can be modeled by an effective spin-1/2 system~\cite{maryasov_anisotropic_2020} with an anisotropic $g$-factor, i.e., a $\mathbf{g}$ tensor, as described by the spin Hamiltonian $H=\mu_B \vec{B} \mathbf{g}\vec{S}$. Here, $\mu_B$ is Bohr's magneton and $\vec{S}$ is the spin vector that contains the three Pauli operators $\vec{S}=\frac{1}{2}(\hat{\sigma}_x, \hat{\sigma}_y, \hat{\sigma}_z)$.

The energy-level splitting in a given magnetic field thus depends on the $\mathbf{g}$ tensor, which varies widely depending on the integration site and its CF levels. The $\mathbf{g}$ tensor can be very anisotropic, with individual principal values exceeding 16 (i.e., eight times that of free electrons) for some sites in silicon~\cite{vinh_photonic_2009, holzapfel_characterization_2025}. This is among the largest effective $g$ factors in any material, which can entail a rather large sensitivity of the energy levels to magnetic fields. Although this might be useful for magnetic field sensing, it can also degrade the spectral stability of optical and spin transitions when subjected to magnetic noise.

As silicon has a low abundance of nuclear spins, one still expects and observes a spin coherence of a few tens of $\si{\micro\second}$ for the electronic spin in materials with natural isotopic abundance~\cite{gritsch_optical_2025}. In purified $^{28}\text{Si}$ hosts, milliseconds of coherence for the electronic spin can be observed ~\cite{berkman_millisecond_2023}. Further improvement is likely possible based on dynamical decoupling~\cite{suter_colloquium_2016}, extending the coherence to the lifetime limit, which can reach the second timescale~\cite{gritsch_optical_2025} and is limited by the spin-lattice interaction and flip-flop processes with paramagnetic impurities and the nuclear spin bath~\cite{wolfowicz_quantum_2021}. Thus, the lifetime depends on the host crystal purity, the temperature, and the magnetic field. At cryogenic temperatures, the Orbach and Raman processes are strongly suppressed, and the lifetime is limited by the direct process and by flip-flop interactions for large and small magnetic fields, respectively.

The large magnetic field sensitivity of optical and/or spin transitions can be circumvented by applying the magnetic field in a specific direction and strength, at which the transitions lose their $B$-field sensitivity~\cite {mcauslan_reducing_2012, yang_zeeman_2022, ortu_simultaneous_2018}. These zero-first-order-Zeeman-shift (ZEFOZ) points can be engineered via the coupling of the electronic spin or Er to the nuclear spin of the isotope $^{167}\text{Er}$, which has a natural abundance of $23\si{\percent}$. 

Even in the absence of precisely aligned fields, the eight hyperfine ground-state levels of the $7/2$ nuclear spin (shown at the right side of Fig.~\ref{fig:erbium_levelScheme}) can enable second-long storage of quantum information, as demonstrated in other host materials~\cite{rancic_coherence_2018}. In silicon, the greater separation of the crystal fields and the lower concentration of magnetic noise sources may even enable coherence times much longer than in current experiments, paving the way for spin-photon interfaces with long-term quantum memory. To this end, the spin state needs to be initialized, coherently controlled, and read out. This can be achieved in nanophotonic resonators, as described in Sec.~\ref{Subsec:NanophotonicDevices}.

\subsubsection{Erbium integration into the silicon host crystal}
Albeit the crystal field is only a perturbation to the free-ion Hamiltonian, it will still determine the precise frequency of the optical transitions of individual erbium dopants. To enable photon-interference experiments that involve different, potentially spatially separated emitters, all dopants should emit within a narrow frequency window that can be bridged either by optical modulators, spectral tuning via strain or Stark effect, or by time-resolved detection~\cite{reiserer_colloquium_2022}. This requires erbium to be reproducibly incorporated into a well-defined lattice site with high yield.

As silicon is a monatomic, highly symmetric crystal, one would not expect single erbium atoms to be incorporated in many different ways. However, it turns out that erbium is gettering in silicon, which means that it tends to form complexes or clusters with other impurities, including other erbium ions. The reason is that erbium prefers ionization to the triply ionized state, which cannot isoelectrically substitute for Si atoms in their covalently bonded tetrahedral lattice positions. Furthermore, there is a significant size mismatch between erbium and silicon, leading to significant lattice deformations around the emitters. Thus, the formation of clusters can be energetically favorable.

For these reasons, it was proven difficult to incorporate erbium into silicon during growth from the melt~\cite{kenyon_erbium_2005}. While equilibrium concentrations around $\sim 10^{18}\si{\centi\meter^{-3}}$ have been achieved, the fraction of erbium that has been integrated into a single site in these experiments is undisclosed owing to limitations in the measurement technique, off-resonant excitation, which will be further discussed in~\ref{subsubsec:OffResExc}. Even if a high concentration of erbium is achieved in such structures, it may still cluster together with other dopants or impurities, harming its optical properties --- in particular, the coherence.

As an alternative to reduce clustering, erbium can be incorporated into silicon using non-equilibrium techniques, such as molecular-beam epitaxy (MBE), chemical vapor deposition (CVD), or ion implantation. An integration yield exceeding $1\,\%$ in the same site has been achieved both by MBE~\cite{vinh_microscopic_2003, vinh_photonic_2009} and by ion implantation~\cite{gritsch_narrow_2022}.

In the latter, the crystal damage that stems from the bombardment with heavy ions needs to be repaired; otherwise, the optical loss of photonic components is prohibitive~\cite{rinner_erbium_2023}. To this end, the sample is subjected to thermal annealing. Typical experiments anneal at temperatures of $\gtrsim\SI{700}{\celsius}$ for several minutes~\cite{kenyon_erbium_2005, yin_optical_2013, przybylinska_optically_1996, priolo_excitation_1998, weiss_erbium_2021, berkman_observing_2023}. At this temperature, most impurities in silicon are mobile~\cite{hull_properties_1999}, such that they can diffuse to and cluster with the erbium dopants. Thus, a large number of erbium sites, each with different spin and optical properties, are found in such samples. In general, one expects that the higher the implantation dose, annealing temperature, and impurity concentration of the starting material, the more distinct clusters can form and the more optical lines will be observed in spectroscopy. While the erbium in these sites will be surrounded by different arrangements of atoms and vacancies, experiments based on Rutherford back-scattering~\cite{kozanecki_lattice_1996} and channeling~\cite{wahl_direct_1997} still indicate that most erbium atoms are integrated in hexagonal or tetrahedral interstitial positions, which are the largest voids in the silicon lattice.

As quantum applications require dopant integration into reproducible lattice sites with a high yield, later experiments focused on using pure starting materials and avoiding annealing above $\SI{600}{\celsius}$. This led to the observation of fewer sites, reproducibly generated with high yield in Cz, FZ, and CVD-grown samples~\cite{gritsch_narrow_2022}. Similar results were even achieved in samples that were commercially processed on a multi-project wafer in a photonic foundry~\cite{rinner_erbium_2023}, demonstrating the potential for up-scaling using established semiconductor manufacturing technology. These sites were called "A" and "B" in~\cite{gritsch_narrow_2022} and subsequently analyzed in detail, as described in Sec. \ref{subsubsec:CF_characterization}. In these works, the erbium concentration was kept low, around $10^{17}\si{\centi\meter^{-3}}$ or lower. This minimizes crystal damage and ion-ion interactions. Still, the concentration is high enough to be achieved with standard ion-implantation tools, and nanophotonic structures with subwavelength-scale mode volumes will, on average, contain one or more emitters at these reproducible sites. 

\subsubsection{Off-resonant and electrical excitation} \label{subsubsec:OffResExc}

Er:Si has only recently been explored in the context of quantum information processing. However, many pioneering experiments had already been conducted, starting in the 1990s. Extensive reviews are provided in Kenyon et al.~\cite{kenyon_erbium_2005}, and Vinh et al.~\cite{vinh_photonic_2009}. The focus of these works was the implementation of on-chip lasers and amplifiers, which eventually turned out difficult due to the mentioned fluorescence quench at elevated temperatures, preventing the realization of sufficient gain in devices at ambient conditions. In the following, we provide a brief summary of these works in the context of quantum applications.

A common feature of all these early experiments is their measurement technique: As the main goal was an electrically pumped gain medium, the works relied on excitation of the erbium by inducing excitons in the silicon layer, either electrically~\cite{zheng_room-temperature_1994} or by off-resonant optical excitation~\cite{przybylinska_optically_1996, priolo_excitation_1998} using a green or blue laser. Subsequently, the emission spectrum was analyzed, typically using a Fourier-transform spectrometer. These studies had two key findings: First, erbium incorporates in many different lattice sites, leading to the observation of inhomogeneous lines with an extent of several $\si{\tera\hertz}$, which would be far too large for quantum applications. Second, most of the erbium did not exhibit any fluorescence upon off-resonant excitation, which was often referred to as the dopants not being "optically active".

While many studies attributed this to rapid non-radiative decays~\cite{kenyon_erbium_2005}, later studies using resonant excitation instead found that a large fraction of the erbium is integrated in sites that cannot be optically excited by excitons~\cite{gritsch_narrow_2022}. Thus, care must be taken when extrapolating results from the early literature to quantum devices --- one must always keep in mind that experiments using off-resonant excitation measure only a small subset of emitters that can capture excitons and then decay radiatively. As defect states are believed to play a key role in the excitation process, this subset will likely exhibit worse coherence than Er:Si on isolated lattice sites, which thus seems better suited for quantum applications.

\subsubsection{Resonant excitation}

To perform spectroscopy on sites of Er:Si, which are decoupled from the conduction band, one needs to perform resonant spectroscopy. However, this is not straightforward in devices fabricated by non-equilibrium techniques, such as ion implantation. Naturally, these structures will contain erbium only in a thin layer $\lesssim \SI{2}{\micro\meter}$ below the surface. Performing absorption spectroscopy perpendicular to this surface is thus difficult, as the long lifetime of rare-earth emitters entails a small oscillator strength~\cite{thiel_rare-earth-doped_2011} and thus a very weak absorption of thin layers. 

The work of Weiss et al.~\cite{weiss_erbium_2021} overcame this limitation by fabricating photonic waveguides into the device layer of a silicon-on-insulator chip, thus increasing the number of erbium emitters that can be excited at a given laser frequency. The work established pulsed resonant fluorescence spectroscopy, in which the emitters are excited by laser pulses typically a few $\si{\micro\second}$ in duration, and the fluorescence is detected with single-photon detectors after the pulses are turned off. In this way, all erbium dopants that decay radiatively can be probed, enabling new insights into Er:Si.

A key finding was that a large fraction of the emitters can be integrated into well-defined lattice sites with a narrow, inhomogeneous distribution of less than $\SI{1}{\giga\hertz}$. This paved the way for quantum applications of Er:Si. Later studies using the same technique~\cite{gritsch_narrow_2022} or in-situ fabricated single-photon detectors~\cite{berkman_observing_2023} revealed very narrow homogeneous linewidths in several lattice sites, down to $\SI{10}{\kilo\hertz}$, paving the way for coherent light emission.

\subsubsection{Crystal field levels, optical and spin properties of site A and B} \label{subsubsec:CF_characterization}

\begin{figure}[ht!] 
\includegraphics[width=1.0\columnwidth]{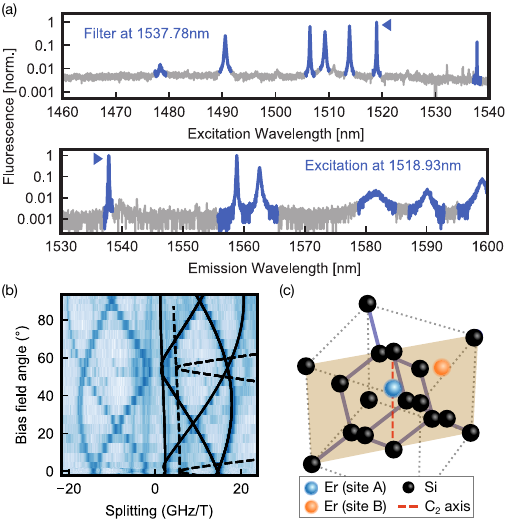}
\caption{     \label{fig:erbium_Properties_Site_A_and_B}
\textbf{Crystal field and magnetic properties of Er:Si in site~A}. (a) Crystal field spectroscopy of site A at low temperature (\SI{2}{\kelvin}). (top) A narrow filter is set to the $Y_1\rightarrow Z_1$ transition, and the excitation laser is swept to measure the excited-state crystal-field levels. (bottom) The excitation laser is fixed at the $Z_1 \rightarrow Y_2$ transition, and the narrow filter is swept, revealing the emission to the lowest six ground-state levels. Adapted from Gritsch et al.~\cite{gritsch_narrow_2022}, Physical Review X, 12, 041009, 2022; licensed under a Creative Commons Attribution (CC BY) license. (b) Spin properties. The sample is rotated in a magnetic field of \SI{1.9}{\tesla} such that the field angle varies from $[001]$ to $[1\bar{1}0]$ and the spin-preserving and spin-flip transitions are measured. The large number of lines indicates low site symmetry, which is determined to be $C_{2v}$ from a fit (solid and dashed lines, shown only for one half of the symmetric spectrum) to the data. (c) Sketch of the silicon unit cell with exemplary positions of the Er dopants according to the extracted site symmetries of $C_{2v}$ (site A) and $C_{s}$ (site B). Thus, site A is located on the two-fold rotational axis (red dotted line) that passes through the unit cell center, and site B on the $\{110\}$ mirror plane (shaded area). Adapted from Holz\"apfel et al.~\cite{holzapfel_characterization_2025}, Advanced Quantum Technologies, 8, 2400342, 2025; licensed under a Creative Commons Attribution (CC BY) license.}
\end{figure}

Even in resonant spectroscopy on the purest samples implanted at moderate temperatures, erbium was found to integrate into several sites, including the previously mentioned sites A and B. These two contain a fraction of $\gtrsim \SI{1}{\percent}$ of the emitters in high-purity starting material, and are particularly promising for quantum applications. First, they exhibit a narrow inhomogeneous broadening in high-purity starting material, $\lesssim \SI{0.4}{\giga\hertz}$, which can be easily bridged by optical modulators. Second, they feature a relatively short radiative lifetime, $\SI{0.14}{\milli\second}$ and $\SI{0.19}{\milli\second}$, respectively, in bulk crystals -- the shortest observed in any erbium host crystal up to now, which can in part be attributed to the large refractive index of silicon~\cite{devries_point_1998}. Third, the emitters exhibit a narrow homogeneous linewidth of $\sim\SI{10}{\kilo\hertz}$, which is only an order of magnitude from being lifetime-limited. Finally, the narrow linewidth is preserved up to $\SI{8}{\kelvin}$, a temperature range that is conveniently accessible by $^4\text{He}$ cryocoolers. The large operating temperature indicates that the emitters are integrated into low-symmetry sites with a large CF splitting $\Delta E_\text{CF}$, which avoids decoherence via the Orbach process whose decay rate is $\propto \exp\left(\Delta E_\text{CF}/k_B T\right)$. Here, $k_B$ is Boltzmann's constant and $T$ is the temperature.

The precise locations of the CF levels have been determined by optical spectroscopy. To this end, a narrowband optical filter was inserted in front of the single-photon detectors in pulsed resonant fluorescence experiments~\cite{gritsch_narrow_2022, holzapfel_characterization_2025}. A typical such measurement is shown in Fig.~\ref{fig:erbium_Properties_Site_A_and_B}(a). In the upper panel, the excitation laser was swept to measure the excited-state CF levels; in the lower panel, the ground-state CF levels were determined by sweeping the filter transmission wavelength. Based on the measured energies, the CF Hamiltonian of the erbium dopants in site A could be reconstructed~\cite{holzapfel_characterization_2025}.

In addition, measurements of the spin-flip and spin-preserving optical transitions while rotating the direction of an external magnetic field allowed a full reconstruction of the $\mathbf{g}$-tensor of site A as depicted in Fig.~\ref{fig:erbium_Properties_Site_A_and_B}(b). Its $C_{2v}$ symmetry was mapped to positions in the Si unit cell that are shifted along the (100) direction from the center of the tetrahedral interstitial position~\cite{holzapfel_characterization_2025}. Site B has an even lower symmetry. Possible positions of both sites are indicated in Fig.~\ref{fig:erbium_Properties_Site_A_and_B}(c). Both sites are polar, such that a non-zero linear Stark coefficient is expected. This can enable emitter tuning by electric fields, but also makes the transition frequency sensitive to charge noise and thus lower optical coherence. Future work will thus aim at quantifying the Stark coefficient, and potentially exploring other sites in Er:Si that may exhibit a non-polar symmetry; so far, however, such sites have only been identified in other host materials~\cite{stevenson_erbium-implanted_2022}.

\subsection{Color centers} \label{subsec:ColorCenters}
Color centers are point defects or atomic impurities in solid-state crystals, which can absorb and emit light. These impurities can consist of a vacancy, an additional atom of the crystal or another element, or a combination of several of these. The physical properties of color centers resemble those of atoms and molecules, with energy levels formed by orbital and spin degrees of freedom. While all color centers emit light and can thus be used as single-photon sources, only some of them exhibit coherent spin states that enable the implementation of spin-photon interfaces.

While early experiments in this direction used diamond as a host material, silicon color centers have recently attracted significant attention due to their telecom operation wavelengths and compatibility with silicon nanofabrication, integrated photonics, and microelectronics. The first experimental studies date back to the 1980s, and in the last decades, an extensive list of color centers has been comprehensively investigated using numerous experimental methods~\cite{bean_electron_1970, davies_optical_1989}. A subset of these centers exhibits a relatively strong luminescence, paving the way for experiments on single-photon emitters~\cite{chartrand_highly_2018, zhang_material_2020, bergeron_silicon-integrated_2020}. 

While most of the observed color centers remain unexplored~\cite{durand_broad_2021}, the so-called T, G, W, and C centers stand out, as they have been studied in more detail and their ZPL falls into the telecom bands, as shown in Fig.~\ref{fig:overview}(c) and Table~\ref{tab:colorcenters}. 
While the W center is a complex of three self-interstitial atoms, the other centers contain carbon, hydrogen, or oxygen impurities (see Fig.~\ref{fig:overview}(a)). 
To create the mentioned color centers, silicon is irradiated by electrons~\cite{bean_electron_1970}, ions~\cite{macquarrie_generating_2021, zhiyenbayev_scalable_2023,hollenbach_wafer-scale_2022}, or intense laser pulses~\cite{jhuria_programmable_2024, quard_femtosecondlaserinduced_2024, gu_end--end_2025} followed by annealing to form the emitters and to recrystallize the material. 
Alternatively, color centers can be formed by molecular beam epitaxy growth of Si with added dopants during growth~\cite{aberl_all-epitaxial_2024}. 
The basic properties of these color centers and their formation will be given in the following sections. 

\begin{table}[htb]
\begin{ruledtabular}
    \begin{tabular}{c|rrrrrr}
        center &  ZPL (nm / meV) & DWF & BE & $\tau$ (ns) & $\eta_{QE}$ & $S$  \\
        \hline
         T (\textsuperscript{1}H) & 1326 / 935  & 0.23  & yes & $\sim$1000 & $\sim$0.2* & 1/2, GS \\
         T (\textsuperscript{2}H) & 1325 / 936  & 0.23  & yes & $\sim$4800 & $\sim$1* & 1/2, GS \\
         W & 1218 / 1018 & 0.40 & yes & 3 to 30 & 0.65  & 0 \\
         C & 1571 / 789 & 0.12*& yes & $\sim$2300& ? & 1, MS \\
         G & 1278 / 970 & 0.15& no  & $\sim$6 & $<$0.18  & 1, MS \\ 
         $\text{G}^*$ & 1278 / 970 & ? & ?  & 30 & >0.5 & ? \\\hline
         $\text{Er}^{3+}$ & 1538 / 806 & 0.23** & & $142\cdot 10^3$ & 1 & 1/2, GS
    \end{tabular}
    \caption{\label{tab:colorcenters}Basic optical and spin properties of selected color centers (top) in silicon, compared to erbium dopants in site A (bottom). ZPL: zero-phonon-line; DWF: Debye-Waller factor; BE: bound exciton excited state; $\tau$: optical lifetime; $\eta_{QE}$: radiative quantum efficiency; $S$ spin state in the ground state (GS) or metastable state (MS). * predicted. **Branching ratio of crystal-field transitions; no phonon sideband.}
\end{ruledtabular}
\end{table}

\subsubsection{T center}

\begin{figure}[ht!] 
\includegraphics[width=1.0\columnwidth]{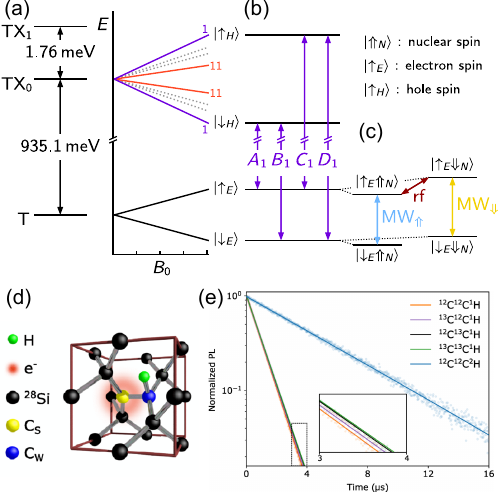}
\caption{     \label{fig:TCenter}
\textbf{Basic properties of the T center.} (a) Energy levels of the ground- and the two BE excited states, (b) which split under a magnetic field, giving rise to four main optical transitions (A-D) associated with the electron and hole spins, and (c) hyperfine transitions associated with nuclear spins. (d) Atomic configuration, showing the arrangement of the two C atoms and the H atom forming the defect. (e) Excited state lifetime measurements of the TX$_0$ transition of different isotopic configurations show a five-fold increase for deuterated T centers. Images adapted from Bergeron et al.~\cite{bergeron_silicon-integrated_2020}, PRX Quantum, 1, 020301, 2020; licensed under a Creative Commons Attribution (CC BY) license, and from Kazemi et al.~\cite{kazemi_giant_2026}, Physical Review Letters, 136, 053602, 2026; Copyright (2026) by the
American Physical Society.}
\end{figure}

First observed~\cite{minaev_thermallyinduced_1981} in 1981, the T center is formed by two carbon and one hydrogen atom with an unpaired electron in a substitutional lattice site~\cite{safonov_hydrogen_1993, lightowlers_hydrogenrelated_1994} (Fig.~\ref{fig:TCenter}a).
Structurally, the pair of carbon atoms is bound within a single lattice site (C-C dimer substituting a Si atom), and the hydrogen atom is bonded to one of the carbons~\cite{dhaliah_first-principles_2022, xiong_computationally_2024}, as shown in Fig.~\ref{fig:TCenter}d. 

The T center features a C$_{1h}$ symmetry, which was initially wrongly assigned to C$_{2v}$~\cite{irion_defect_1985} in 1985, later corrected~\cite{safonov_carbon-hydrogen_1995} in 1994, and recently confirmed using the previous strain- and new electric field perturbation measurements of its optical transitions~\cite{clear_optical-transition_2024}. This rather low symmetry results in 24 defect orientations and a non-zero linear Stark effect~\cite{clear_optical-transition_2024}.
    
The ZPL of the T center is at 1326 nm (935 meV) in the telecom O-band. The optical emission originates from the decay of a bound exciton (BE) from a delocalized valence-band state to a localized defect state~\cite{dhaliah_first-principles_2022} (Fig.~\ref{fig:TCenter}b).
In the BE excited state, the two electrons form a singlet, and the hole states split into two doublets labelled TX0 and TX1. The lifetime of the excited state of T centers is rather long for color centers, around~\cite{dhaliah_first-principles_2022, veetil_enhanced_2025} $\sim 1~\mu\text{s}$. The optical lifetime is even 5 times longer for T centers that contain deuterium rather than hydrogen, revealing a large isotope effect in the optical properties~\cite{kazemi_giant_2026}.

The quantum efficiency of the T center is still under debate, with extracted values ranging from >23.4\%~\cite{johnston_cavity-coupled_2024} to estimates of 18.1\% for hydrogen-based emitters. For deuterated T centers, an improved quantum efficiency, near unity, is predicted~\cite{kazemi_giant_2026}, suggesting that this configuration may be beneficial for quantum applications. In both configurations, the T center exhibits a DWF~\cite{bergeron_silicon-integrated_2020} of 0.23. The inhomogeneous broadening has been measured at 37~GHz in implanted natural SOI~\cite{higginbottom_optical_2022}. Compared to the 33~MHz reported for isotopically purified bulk $^{28}$Si crystals~\cite{bergeron_silicon-integrated_2020}, the broadening is due to strain gradients intrinsic to SOI fabrication and to implantation-induced or intrinsic defects.

In bulk, isotopically purified silicon, homogeneous linewidths of 690(10)~kHz were observed, exceeding the theoretical lifetime limit of 170~kHz for the T center only about fourfold \cite{deabreu_waveguide-integrated_2023}. Instead, measurements in nanostructures and SOI found orders of magnitude larger values, ranging from 1 to 10~GHz~\cite{higginbottom_optical_2022, komza_multiplexed_2025}. Recent experiments indicate that the resonant excitation laser is the main cause of spectral diffusion, reaching linewidths of 110~MHz over a few ms when using resonance check techniques~\cite{bowness_laser-induced_2025, zhang_laser-induced_2025}. The spectral wandering is thus attributed to charge fluctuations at nearby defects and surfaces that cause a Stark shift due to the center's low C$_{1h}$ symmetry and are induced by laser pulses. Besides charge noise, a significant phononic contribution is observed via the temperature dependence of the linewidth, saturating~\cite{bergeron_silicon-integrated_2020} below 1.5~K.

The T center exhibits a spin-$\tfrac{1}{2}$ ground state and an allowed spin-conserving optical transition~\cite{higginbottom_optical_2022}, making it very promising as a spin-photon interface (Fig.~\ref{fig:TCenter}c). The hyperfine interaction couples the electronic spin to the nuclear spin of the hydrogen atom. In isotopically purified $^{28}$Si bulk crystals, this has enabled the measurement of coherence times of 2.1~ms and 1.1~s for electron and nuclear spin, respectively~\cite{bergeron_silicon-integrated_2020}. In a thin film photonic device, spin echo coherence times of 0.4~ms, 112~ms, and 67~ms have been reported for electron, hydrogen, and proximal $^{29}$Si spins. This has enabled the generation of nuclear spin entanglement with a fidelity~\cite{song_entanglement_2025} of 0.77. Recently, a route towards utilizing the optical interface with reduced decoherence of the hydrogen spin has been proposed~\cite{brunelle_silicon_2025}.

To create T centers in nanophotonic devices, nearly all reported works follow slight variations of the fabrication process described in Ref.~\cite{macquarrie_generating_2021}. 
The process consists of two stages of ion implantation and annealing. First, C is implanted, and the crystal is annealed at 900 to $\SI{1000}{\celsius}$ to heal the lattice. This is followed by an H implantation step often with a similar dose to the initial C implantation, and a second annealing step at 400 to $\SI{450}{\celsius}$. This process, especially the second step, may pose stringent requirements on the annealing temperature and time. 
While generating T centers with low densities has been widely reported, achieving high concentrations is expected to be challenging due to the tri-atomic nature of the defect~\cite{dhaliah_first-principles_2022, macquarrie_generating_2021}, which may limit applications requiring high-density ensembles such as amplification and lasing, photonic quantum memory, and transduction~\cite{higginbottom_memory_2023, khalifa_robust_2025, rinner_quantum_2026}.

After creating T centers, lithography and etching are used to fabricate nanophotonic structures; these processes may introduce additional defects leading to an undesired fluorescence background. Thus, some groups also perform the second annealing step after etching~\cite{komza_multiplexed_2025}. Recent experiments performed on nanoscale waveguides and resonators~\cite{lee_high-efficiency_2023, deabreu_waveguide-integrated_2023, johnston_cavity-coupled_2024, photonic_inc_distributed_2024, komza_multiplexed_2025, day_probing_2025, dobinson_electrically_2025} will be discussed in depth in Sec.~\ref{sec:Emitters_in_nanostructures}.

\subsubsection{G center}
\begin{figure}[tb] 
\includegraphics[width=1.0\columnwidth]{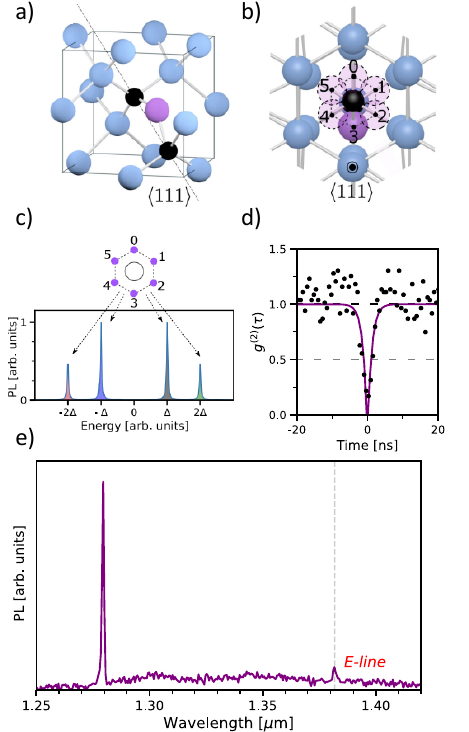}
\caption{ \textbf{Basic properties of the G center.} (a) The atomic configuration consists of two substitutional carbon atoms (black) and an interstitial silicon atom (purple). (b) The silicon atom can occupy different rotational degrees of freedom when seen along the defect axis. (c) The different Si positions split the energy levels into four lines in unstrained crystals, which can be observed in the ZPL PL spectrum. (d) Second-order autocorrelation measurement performed on photons from a single G center, showing a fit to a two-level system. (e) A PL spectrum from a single G center showing its ZPL and its characteristic E-line local vibrational mode. Images adapted from Durand et al.~\cite{durand_hopping_2024}, Physical Review X, 14, 041071, 2024; licensed under a Creative Commons Attribution (CC BY) license.\label{fig:gcenter}
}
\end{figure}

The G center is formed by two carbon and a silicon atom~\cite{song_bistable_1990}, and can adopt multiple configurations. While two of its configurations, called A and C, are optically dark~\cite{deak_kinetics_2023}, the B configuration exhibits photoluminescence. 
This configuration corresponds to a neutral ground state in which two carbon atoms occupy substitutional sites and are bridged by a silicon interstitial (Fig.~\ref{fig:gcenter}a). 
Like the T center, the G center features C$_{1h}$ symmetry~\cite{brower_epr_1974}. 
A unique property of the G center is the rotational degree of freedom of the interstitial silicon atom, as shown in Fig.~\ref{fig:gcenter}b.

The transition dipole of the G center is oriented along the defect’s principal axis, as confirmed experimentally by observing its polarization axis along either [110] or [1$\bar{1}$0] in confocal measurements using (100) plane samples. Due to its rotational degree of freedom, the emission of the G center is not purely dipolar, but shows a limited visibility~\cite{durand_genuine_2024} of 62\%. The ZPL of the G center is at 1278~nm (970~meV) in the telecom O-band, and was first observed~\cite{watkins_epr_1976} in 1976. 
The rotational degree of freedom of the defect splits the ZPL into six spectral lines. 
Without strain, two of the lines are degenerate, leading to a 4-peak spectrum with double intensity in the central peaks~\cite{chartrand_highly_2018, durand_hopping_2024, udvarhelyi_identification_2021}, see Fig.~\ref{fig:gcenter}c. An additional distinguishing spectral feature of the G center is the E-line, a PSB resonance arising from a local vibrational mode of its carbon pair~\cite{baron_single_2022}, see Fig.~\ref{fig:gcenter}e.

The G center excited-state lifetime is short compared to the T center, below 6~ns for its spin singlet ZPL transition~\cite{durand_genuine_2024, prabhu_individually_2023, komza_indistinguishable_2024, saggio_cavity-enhanced_2024}. Its quantum efficiency is low, upper bounded~\cite{saggio_cavity-enhanced_2024} by 18\%, and its DWF is~\cite{beaufils_optical_2018} 15\%. While most measurements show stable operation, above-band optical irradiation experiments with powers an order of magnitude above saturation show significant spectral shifts and deactivation of single G centers~\cite{prabhu_individually_2023}.

In SOI photonic waveguides, the inhomogeneous broadening of the G center is below 200~GHz~\cite{prabhu_individually_2023, durand_genuine_2024}. The homogeneous broadening upon resonant excitation still needs to be determined. With off-resonant excitation, 2.8~GHz linewidths have been reported for long timescales, which reduces to 0.4~GHz at 25~ns in two-photon interference measurements performed on photons consecutively emitted by a G center in a SOI waveguide~\cite{komza_indistinguishable_2024}.

While the G center’s neutral ground state is a closed-shell singlet and thus has no accessible spin, its excited-state manifold includes an optically bright singlet and a slightly higher-lying, optically dark, metastable triplet state (678~meV above the ground state)~\cite{deak_kinetics_2023, udvarhelyi_identification_2021}. This was first reported in the 80s through optically-detected magnetic resonance (ODMR) measurements of ensembles~\cite{brower_epr_1974, lee_optical_1982}, and recently demonstrated in single G centers, including their coherent spin control~\cite{cache_single_2025}. The observed triplet state could be used as a quantum memory; however, the storage time would be limited by the yet unknown optical lifetime of the metastable state. It has been proposed that longer timescales might be accessible via a transfer of the quantum information to a long-lived nuclear spin of the defect atoms or the lattice~\cite{lee_readout_2013}.

Several procedures have been used to create G centers in silicon. As with other carbon-related color centers in silicon, a common fabrication technique involves a carbon implantation step followed by rapid thermal annealing. Non-equilibrium processes in this procedure are hypothesized to reduce the G center density down to the level of single emitters~\cite{prabhu_individually_2023, saggio_cavity-enhanced_2024, zhiyenbayev_scalable_2023}. Some experiments also used proton irradiation~\cite{baron_single_2022, durand_genuine_2024} or observed G center formation after dry etching of C-doped silicon~\cite{durand_genuine_2024}. As this is a common step in nanofabrication, it has been hypothesized that the G centers observed in earlier works may have been created by the etching process rather than by the combination of implantation and annealing~\cite{cache_single_2025}. Alternative methods have recently been reported for localized formation of G centers using focused silicon-ion beams~\cite{hollenbach_wafer-scale_2022} and optical irradiation~\cite{jhuria_programmable_2024, quard_femtosecondlaserinduced_2024, gu_end--end_2025}, including formation in pre-patterned nanophotonic cavities~\cite{gu_end--end_2025}.

Recent single-defect spectroscopy~\cite{durand_genuine_2024} has shown that a substantial fraction of telecom-band emitters previously attributed to the G center actually belong to a distinct defect family, denoted as G$^{*}$, which also exhibits a ZPL near 1.28~$\mu$m but with a much broader inhomogeneous distribution. This confusion was first hypothesized~\cite{baron_single_2022, saggio_cavity-enhanced_2024} and recently confirmed experimentally~\cite{durand_genuine_2024}. There are several differences between the emitters that allow a clear discrimination: first, G$^{*}$ spectra do not display the characteristic $\sim$1382~nm E-line local vibrational mode replica of G centers~\cite{durand_genuine_2024}. Second, the G$^{*}$ center has a longer excited state lifetime~\cite{redjem_single_2020} of 30~ns, a higher quantum efficiency~\cite{redjem_single_2020, redjem_all-silicon_2023, durand_genuine_2024} above 50\%, and a broader inhomogeneous linewidth~\cite{redjem_single_2020, durand_genuine_2024} of 12~nm.

G$^{*}$ also features a purely dipolar emission with broad angular variability that skips the C$_{1h}$ principal axes that would be expected for the G center~\cite{redjem_single_2020}. Open questions include the G$^{*}$ center atomic structure, its point group, its homogeneous broadening, and its spin properties~\cite{durand_genuine_2024}. In addition, a defect named G' has been recently observed in MBE-grown samples, which has the same PL lineshape as that of the G center but red shifted by about 17~meV~\cite{aberl_all-epitaxial_2024}. While the theory predicts properties similar to those of the G center, such as the availability of an optically readable metastable triplet state, experimental confirmation of most of these properties is needed.

Due to their relative ease of fabrication and isolation, the G- and G$^{*}$ centers have been used as a testbed in many experiments on isolated silicon color centers. These experiments include the first observation of isolated color centers in silicon (G$^{*}$)~\cite{redjem_single_2020, hollenbach_engineering_2020}, the demonstration of waveguide- (G)~\cite{prabhu_individually_2023} and cavity integration (G$^{*}$ and G)~\cite{redjem_all-silicon_2023, saggio_cavity-enhanced_2024}, as well as strain tuning and nanoscale localization (G)~\cite{buzzi_spectral_2025}.

\subsubsection{W center}
Historically observed as the so-called W-line in irradiated or ion-implanted Si, the microscopic structure of the W center remained under debate for decades~\cite{coomer_interstitial_1999, jones_self-interstitial_2002, gharaibeh_molecular-dynamics_1999} until it was confirmed as a silicon tri-interstitial. 
Tight-binding molecular dynamics calculations~\cite{richie_complexity_2004} revealed that an extended trigonal configuration (labelled I3-V) having C$_3$ symmetry satisfied all the experimentally observed properties, while advanced density functional theory (DFT) calculations confirmed its superior stability over the other tri-intersitial cluster configurations~\cite{baron_detection_2022}. 
In effect, in the W center, a single defect-derived state is located very close to the valence-band edge. 
In the equilibrium neutral-charge state, this level is occupied by two electrons and resonates with the valence band. 
Upon optical excitation, an electron is promoted to the conduction band (or a conduction-band-like extended state), leaving behind a tightly bound hole in what becomes an in-gap acceptor level. 
The Coulomb attraction between the electron–hole pair forms a bound exciton, which then recombines to give the observed W center photoluminescence. 
Symmetry analysis for I3-V indicates an optical transition with the dipole oriented along ⟨111⟩.

The ZPL line of the W center is around~\cite{davies_optical_1989, tan_ion_2003} 1218~nm (1018 meV). It features a DWF of 40\%~\cite{baron_detection_2022, veetil_enhanced_2025} and a quantum efficiency of 65\%~\cite{lefaucher_purcell_2024}, both larger than those of the G center and the hydrogen (\textsuperscript{1}H) T center. Its dipole orientation matches the [110]/[1$\bar{1}$0] directions expected from the atomic configuration. The excited-state lifetime of the W center has been measured to range between 3~ns to 30~ns for different emitters under different conditions~\cite{buckley_optimization_2020, baron_detection_2022}.

Inhomogeneous broadening measurements range from narrow ensemble linewidths~\cite{chartrand_highly_2018} of 150~MHz, measured in isotopically purified $^{28}$Si bulk crystals, up to 240~GHz in 70~nm-thick SOI~\cite{baron_detection_2022}. Homogeneous broadening measurements have been limited to off-resonant excitation, reporting measurement-limited linewidths below 24~GHz~\cite{baron_detection_2022}.

W centers have been fabricated via silicon ion implantation followed by rapid thermal annealing~\cite{baron_detection_2022} at $\SI{1000}{\celsius}$ for 20~s. Annealing studies suggest that lower temperatures enable higher density formation~\cite{buckley_optimization_2020}. In addition, localized formation has been achieved via focused silicon-ion beam~\cite{hollenbach_wafer-scale_2022}, and femtosecond laser irradiation~\cite{quard_femtosecondlaserinduced_2024, jhuria_programmable_2024}.

The W center has a singlet ground state. Unlike the G- and T centers, it has neither a spin-bearing ground state nor a metastable triplet or other paramagnetic state that would be required for spin-photon interfaces. Therefore, current interest in W centers is largely focused on their use as single-photon emitters~\cite {veetil_enhanced_2025}. To this end, W centers have been integrated into photonic devices, including circular Bragg grating cavities for ensembles~\cite{veetil_enhanced_2025} and single emitters~\cite{lefaucher_purcell_2024}, as well as photonic waveguides and microring resonators~\cite{tait_microring_2020}. In addition, using integration into p-i-n junctions, electroluminescence has been observed~\cite{ebadollahi_fabrication_2024}.

\subsubsection{C center}
The C center is formed by a carbon-oxygen interstitial pair (C$_\text{i}$O$_\text{i}$) and features the same C$_{1h}$ point-group symmetry as the G and T centers~\cite{davies_carbon-related_1989, ayedh_formation_2020}. 
They can be generated by carbon ion implantation and rapid thermal annealing at 1000~C for 20~s, followed by high-energy (1~MeV) proton irradiation~\cite{wen_optical_2025}. Recently, formation via light-ion irradiation has also been reported~\cite {crosta_formation_2025}.

The ZPL emission of the C center occurs in the telecom L band~\cite{davies_model_1987, thonke_carbon_1984, trombetta_identification_1987, wen_optical_2025} around 1571~nm (789~meV). To emit light, the defect captures a hole via a short-range defect potential, and then an electron via a long-range Coulomb potential. This leads to the formation of a hole-attractive isoelectronic bound exciton with a hydrogen-like series of excited-state energy levels. The vibrational spectrum of the C center was identified by local mode spectroscopy (with modes at approximately 64.5, 72.6, 138.1, and 145.3 meV)~\cite{davies_model_1987, thonke_carbon_1984}. Its DWF has been predicted~\cite{udvarhelyi_l-band_2022} to be 12\%, but it still needs to be measured accurately. The excited state lifetime of the lowest exciton transition is comparably long, $\SI{2.3}{\micro\second}$~\cite{bohnert_transient_1993}. Inhomogeneous linewidths down to 45~MHz have been reported for isotopically purified $^{28}$Si bulk crystals~\cite{chartrand_highly_2018}.

While the ground state and most excited states of the C center are singlet-singlet transitions, it also features a metastable triplet excited state predicted by theory and confirmed by experiments~\cite{udvarhelyi_l-band_2022, bohnert_transient_1993, wen_optical_2025}. The presence of the triplet state enables an intersystem crossing (ISC), which has been measured via phosphorescence decay to the ground state at temperatures below 2~K~\cite{ishikawa_optical_2011}. Recent work has demonstrated that the C center's spin system can be accessed optically via ODMR measurements of ensembles with and without a magnetic field~\cite{wen_optical_2025}. Time-resolved PL measurements provided the radiative and ISC lifetimes of 14 and 2.8~\textmu s~\cite{wagner_excitation_1984}, respectively. Further experiments show that the optically bright excited states decay over tens of microseconds, while population transferred via the ISC to the triplet state exhibits much longer non-radiative lifetimes, exceeding $1.4~\mathrm{ms}$ for $m_s = 0$ and $10~\mathrm{ms}$ for $m_s = \pm 1$. This pronounced lifetime contrast enables spin-dependent carrier trapping and thus enables ODMR experiments~\cite{wen_optical_2025}.

\subsection{Theoretical methods} \label{subsec:TheoreticalMethods_NewEmitters}

In the previous sections, we demonstrated that theoretical modeling has played a crucial role in identifying and characterizing color centers in silicon; however, the discussion primarily focused on experimentally determined properties of erbium dopants and color centers. In the following, we provide a detailed description of the theoretical frameworks used to model such defects and show how these approaches enable the systematic search for new color centers with properties that may be advantageous compared to those identified to date.

\subsubsection{Ab initio modeling of color centers}
First-principles calculations, such as Kohn-Sham DFT, are the workhorse for predicting the correct defect properties in materials for quantum applications. For accurate calculations, a sizable part of the crystalline environment must be taken into account. For silicon, this means that accurate calculations typically require a cell of 512 atoms ~\cite{udvarhelyi_identification_2021}. First principles many-body methods for calculating the ground state, as e.g. the random phase approximation (RPA) ~\cite{bohm_collective_1953}, cannot yet be applied to such large systems except if important vertex corrections are included ~\cite{liu_cubic_2016,riemelmoser_optimized_2021}. For the correct simulation of materials for quantum applications, the host material’s band structure should be reproduced accurately, and the defect-induced electronic states must be correctly localized. This ensures the correct calculation of charge transition levels and excited states.

In practice, semilocal DFT (GGA) functionals, such as PBE, significantly underestimate band gaps and over-delocalize defect states. For instance, PBE predicts silicon’s band gap to be $\approx0.7$~eV versus the experimental $1.1$~eV. The fundamental cause is that the exchange energy shows no derivative discontinuity at integer occupation numbers, and is convex between two integer values~\cite{perdew_density-functional_1982, perdew_generalized_1996, yang_more_2016}. As a consequence, both the bandgap and the localization of defect states are severely underestimated. The state-of-the-art solution is to “climb Jacob’s ladder” to hybrid functionals, e.g., the screened hybrid HSE06 functional (Heyd-Scuseria-Ernzerhof, with ~25\% nonlocal Fock exchange) has become the main choice for solid-state defects~\cite{perdew_jacobs_2001, heyd_hybrid_2003, krukau_influence_2006}. Hybrid DFT restores much of the missing derivative discontinuity by mixing in Hartree-Fock exchange, thereby correcting the band gap and improving the localization of defect orbitals. With appropriate tuning (e.g., accounting for dielectric screening in the host), HSE06 can yield defect level positions and band gaps in excellent agreement with many-body $GW$ calculations~\cite{deak_choosing_2017, deak_carbon_2019, han_defect_2017}. In cases where a defect contains strongly localized levels, such as transition-metal $d$ states that experience on-site correlation, a Hubbard-like correction can be added on top of HSE06, commonly referred to as HSE06+$V_w$, to further improve agreement with experiments~\cite{ivady_theoretical_2014}. In cases where the strongly localized defect is a rare-earth ion with $f$ electrons, this is a challenge for conventional DFT and often requires beyond-DFT approaches. One promising strategy is to use wavefunction-based embedded cluster methods, where a finite cluster is treated with high-level many-body techniques such as complete active space self-consistent field (CASSCF)~\cite{olsen_casscf_2011}, second-order M{\o}ller–Plesset perturbation theory~\cite{head-gordon_mp2_1988} and coupled cluster methods~\cite{jeziorski_coupled-cluster_1981}. These methodologies have been applied to investigate optical transitions in oxides~\cite{wen_site_2019}; however, they have not been used for $f$-electron elements in bulk Si yet.

Using a well-parameterized hybrid functional not only provides accurate ground-state geometries but also a reliable spectrum of single-particle states for the defect. Excited-state properties can then be obtained without using very computationally expensive methods, for example, by using the $\Delta$SCF or constrained DFT techniques to construct excited Slater determinants and estimate optical transition energies~\cite{gonze_constrained_2022}. It has been shown that $\Delta$SCF with HSE06 reproduces the results of full $GW$/Bethe-Salpeter equation calculations for defect excitations~\cite{hanke_many-particle_1979}. Moreover, it is crucial to include multiple corrections arising from the supercell approach. For example, for charged defects, techniques like the Freysoldt charge correction~\cite{freysoldt_fully_2009} are used. In slabs, one can choose between the methods of Freysoldt and Neugebauer~\cite{freysoldt_first-principles_2018} and that of Komsa and Pasquarello~\cite{komsa_finite-size_2013}. In case of optical transitions, a correction similar to the charge correction must be applied~\cite{gake_finite-size_2020, falletta_finite-size_2020}. In fact, the interaction between repeated charges influences not only the total energy, but also the potential and, therefore, the one-electron states as well. This influence may turn detrimental in slab calculations (for surface defects). The self-consistent potential correction (SCPC) method takes care of that in bulk and slab models alike~\cite{chagas_da_silva_self-consistent_2021}.

\textit{Ab initio} approaches now enable a complete magneto-optical characterization of colour centers. Specifically, first-principles calculations can predict ZPL energies, PSB lineshapes, fine structure from spin-spin and spin-orbit interactions, and even spin-resonance parameters. Their predictive power was first benchmarked in wide-bandgap semiconductors, such as diamond, where DFT reproduces the NV center's electronic structure, radiative and non-radiative lifetimes, and response to external perturbations~\cite{gali_ab_2019, gali_recent_2023}. In silicon, hybrid-DFT has reached a similar agreement with experiments for the telecom-wavelength emitters discussed above and can resolve their microscopic origins. For the G center, first-principles theory recently unambiguously identified its structure and the ro-vibrational spectrum of the optical excited state coming from the rotating Si interstitial in the complex~\cite{udvarhelyi_identification_2021}. Furthermore, the position of the metastable triplet level was determined together with its spin-Hamiltonian parameters, such as the zero-field splitting and the hyperfine couplings, and this helped understand why ODMR indicates a symmetry change upon thermal activation~\cite{udvarhelyi_identification_2021}. Hybrid functional and constraint DFT likewise describe the bound-exciton properties of T center, predicting a ZPL of 0.985~eV and a microsecond radiative lifetime, close to the experimental findings~\cite{dhaliah_first-principles_2022}. Beyond these established carbon-complexes, recent theory has also identified the single carbon interstitial telecom emitter in Si, confirming its bound exciton recombination in the neutral charge and identifying a metastable triplet that can serve as quantum memory~\cite{deak_quantum_2024}. Importantly, \textit{ab initio} modelling also predicts the impact of external perturbations to the properties of these color centers, for example, in the G center strain changes the symmetry, splits the optical lines, and modifies the selection rules~\cite{durand_hopping_2024}. Comparable progress has been reported in other silicon centers, such as the W center~\cite{baron_detection_2022} and the neutral C center~\cite{udvarhelyi_l-band_2022}, where DFT reproduces the luminescence lineshape and identifies the origin of the 0.79~eV "C-line" as the emission from the C$_i$O$_i$ complex while predicting a stable triplet excited state enabling ODMR-based spin control.

Despite the progress made from hybrid-DFT approaches, further methodological advances will be required to describe color centers in silicon. In particular, color centers such as Er will likely require the use of beyond-DFT frameworks, such as wavefunction-based and embedding methods, to accurately capture multiplet structure and excited-state dynamics. At the same time, the description of spin defect properties and their dependence on strain and electric fields remains to be determined for most of the color centers in Si. Furthermore, understanding the ionisation and non-radiative pathways remains an open challenge that is a critical issue in quantum-optics protocols and in the potential photoelectrical~\cite{Salomon_telecom_2025} readout of defect spins. Addressing these effects theoretically can open the road for the creation of new silicon-based quantum photonic platforms. 

\begin{figure}[tb] 
  \centering
  \includegraphics[width=1.0\columnwidth]{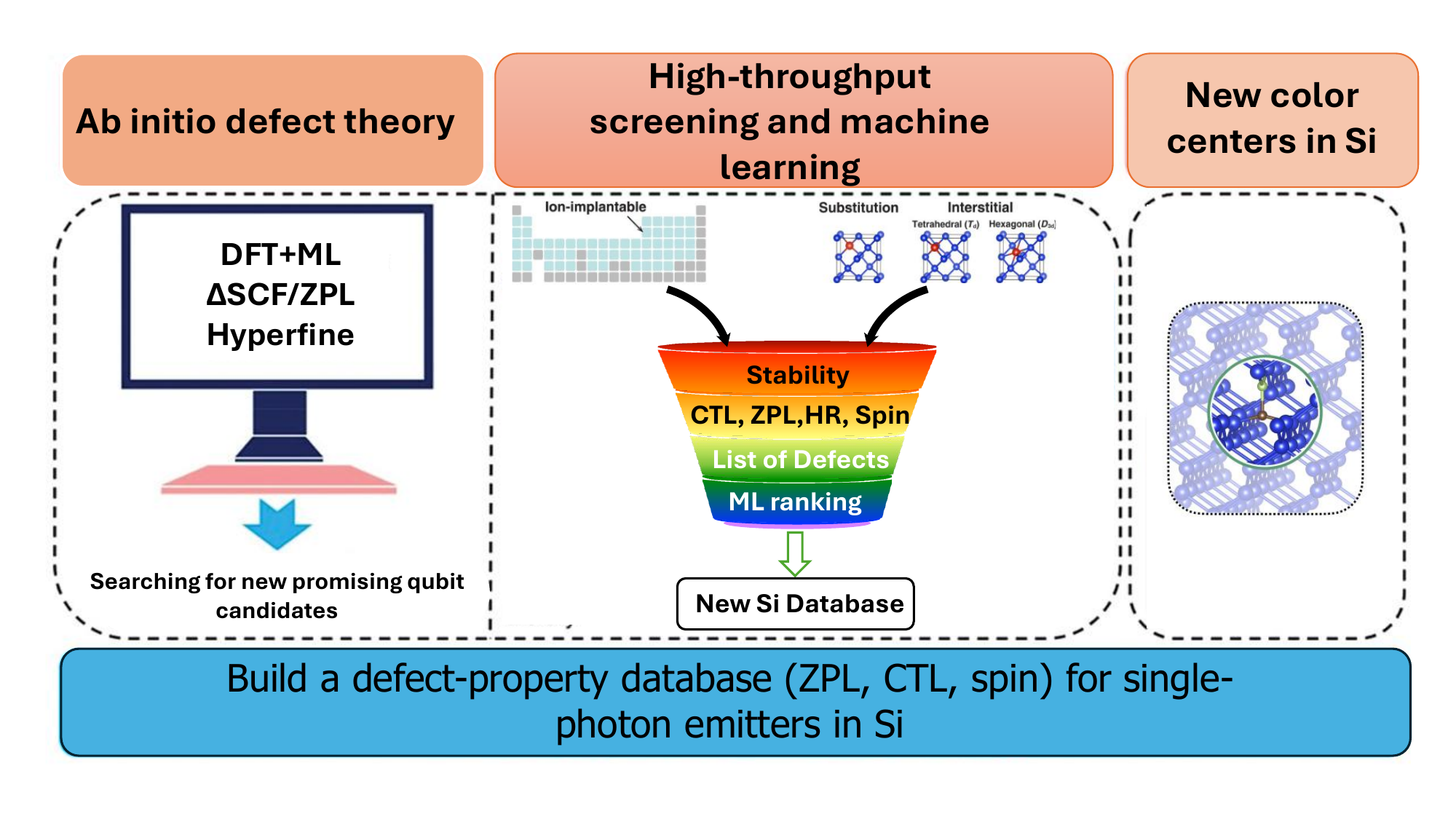}
  \caption{\textbf{Summary of the theory-guided defect-discovery workflow for single-photon emitters in silicon.}
  \textit{Ab initio} theory (left), using density functional theory to compute key parameters assisted by machine learning, is applied to individual defect candidates. These descriptors enable high-throughput screening and machine-learning prioritization (center). Candidate ion-implemented substitutional and interstitial complexes are filtered and ranked to build a silicon defect--property database and identify new
  promising color centers in Si (right). The figure is inspired by~\cite{xiong_high-throughput_2023,shahzad_accelerating_2024}.}
  \label{fig:dft-ml-workflow}
\end{figure}

\subsubsection{Search for new emitters}

Beyond individual point defect calculations, there is a growing trend to use high-throughput computational screening and machine learning to discover new solid-state qubits. In these workflows, large databases of impurity configurations and defect complexes are generated and then filtered to target the quantum defect requirements, such as formation energies, stable charge state, spin multiplicity, and excited-state properties, including transition dipole moment and ZPL. Recent efforts have reached a scale of thousands of defects across multiple hosts, including Si, SiC, and diamond~\cite{davidsson_na_2024, davidsson_discovery_2024, ivanov_database_2023}. Specifically, in the work of Ivanov et al.~\cite{ivanov_database_2023} high-throughput calculations for more than 50{,}000 point defects and complexes were performed in Si, SiC, and diamond, and the associated computed properties and relaxed defect structures are publicly available in a continued expanding database. A similar high-throughput effort has also been performed by Davidsson et al.~\cite{davidsson_na_2024,davidsson_discovery_2024} who used a first principles workflow to identify new spin-qubit candidates, such that a series of “modified $V_{\text{Si}}$” vacancy complexes in 4H-SiC, $NV$ center-like defects in diamond, and atomic clock-like dopant-vacancy centers in oxides. All the computed properties and corresponding structures are available through an open database. Separate databases have also been created for two-dimensional materials~\cite{bertoldo_quantum_2022, cholsuk_hbn_2024}. Such publicly available defect databases accelerate the discovery of new qubits by making large-scale calculations searchable, enabling comparisons and data-driven prioritization of the promising candidates for experimental validation and device integration. 

High-throughput defect-discovery studies typically begin from a small set of experimentally observed centers, such as the NV center in diamond or the T center in silicon, and translate their properties, such as spin multiplicity, well-localized in-gap levels, and target ZPL, into computational filters. Then, a large number of point defects and defect complexes are generated and evaluated against these filters. The most promising candidates are ranked for higher-accuracy calculations and, where possible, for experimental synthesis and characterization routes. This workflow is summarized in Fig.~\ref{fig:dft-ml-workflow}.  
For silicon, high-throughput first-principles screening recently identified a family of T center-like substitutional complexes, $(A\!-\!C)_{\mathrm{Si}}$ with $A=\mathrm{B, Al, Ga, In, Tl}$, which preserve the T center’s valence electron count and are predicted to emit at telecom wavelengths. Several candidates offer improved radiative lifetime or reduced optical linewidth, and feasible formation routes via carbon capture by $A_{\mathrm{Si}}$ or via hydrogenated precursors followed by dehydrogenation were proposed~\cite{xiong_computationally_2024}. Xiong et al~\cite{xiong_high-throughput_2023} also report a first-principles-based screening of more than 1000 charged point defects to identify viable spin-photon interfaces that operate at telecom wavelengths. Using the single-shot hybrid functional (labeled as HSE0) to filter the spin-multiplicity, transition dipole moment, ZPL, and electron-phonon coupling, three interstitial defects, namely Fe$_i^0$, Ti$_i^+$, and Ru$_i^0$, emerged as the most promising candidates. Similar high-throughput screening workflows are now being extended to 2D hosts, for example, in monolayer WS$_2$, a substitutional-defect database enabled identification of the neutral Co$_\mathrm{S}$ center as a promising in-gap defect with a computed telecom transition, which was subsequently fabricated by site-selective STM manipulation and benchmarked by STM/STS against first-principles predictions~\cite{thomas_substitutional_2024}.

However, the accuracy and validation of these big-data approaches remain an active concern. Due to computational cost, most defect databases rely on PBE or single-shot hybrid-level calculations and necessarily make compromises on supercell size or level of theory. As a result, predicted properties may be rather uncertain. Moreover, search algorithms tend to focus on thermodynamically stable defects, but this strategy can overlook complex yet technologically crucial qubits such as the G center in Si~\cite{gali_recent_2023}. In addition, using functionals such as GGA or smaller supercells in high-throughput workflows can yield inaccurate results for defect stability, either by ignoring stable defects or by incorrectly predicting unstable configurations as stable. This limitation is discussed in detail by Huang and Lee~\cite{huang_defect_2012}, who showed that PBE predicts as stable the boron vacancy in -3 charge ( \(\mathrm{V_B^{-3}}\)) in hBN, whereas this stability is not reproduced with HSE06. Meta-GGA functionals can be an alternative approach, as they have demonstrated high accuracy in modeling quantum defects, particularly in 4H-SiC~\cite{abbas_theoretical_2025}. Their performance in predicting defect formation energies closely approaches that of hybrid functional, while they have also been reliable in identifying the defect charge states. In addition, the r$^2$SCAN functional has been shown to predict zero-phonon line energies more accurately than PBE, with smaller errors and a more consistent description of both ground- and excited-state properties. These results highlight meta-GGA methods as a computationally efficient alternative to hybrid functionals for large-scale defect screening.

High-throughput screening of spin defects is connected to machine learning (ML) to prioritize candidate spin-photon interfaces at scale. ML is now being used to speed up major computational bottlenecks in these workflows, such as defect reconstruction, where guided searches can reduce the number of first-principles relaxations. Furthermore, ML interatomic potentials can accelerate the Huang--Rhys and lineshape calculations, making vibronic simulations for photoluminescence sidebands and DWF more efficient. Specifically, Mosquera-Lois et al.~\cite{mosquera-lois_machine-learning_2024} introduced a machine-learning force field to accelerate defect-structure searching, demonstrating its ability to predict the ground state for most neutral point defects while reducing the number of relaxations. The same approach allows more thorough sampling and often finds lower-energy vacancy structures that a DFT-only search can miss, thereby improving both the speed and the reliability of defect geometry determination. For quantum defects in Si, this would be crucial because it could provide the correct ground state of filtered defects, which is essential for identifying promising color centers. In a separate work, Sharma et al.~\cite{sharma_accelerating_2025} presented a framework that accelerates defect photoluminescence calculations by replacing the DFT phonon-mode step with phonons from universal machine-learning interatomic potentials, while having near–DFT accuracy for Huang–Rhys factors and PL lineshapes. Benchmarking on a dataset of 791 color centers shows agreement to full DFT and yields speedups exceeding an order of magnitude, allowing PL spectra to be calculated in high-throughput defect screening. \\
Complete workflows using ML for selecting efficient quantum defects have also been applied. Frey et al.~\cite{frey_machine_2020} present a screening workflow that combines deep transfer learning for host selection with DFT, and ML models to predict defect properties, enabling fast exploration of many possible point defects in 2D materials for quantum emission. Specifically, in this work, they start from nearly 4000 two-dimensional (2D) host structures and use graph-network transfer learning trained on large bulk-crystal datasets to predict which hosts have properties suitable for accommodating defects. Afterwards, they generate nearly 10,000 candidate defect structures across TMDs, hBN, and wide-band-gap 2D materials and compute the band structures and formation energies using DFT for more than 1000 defects to train ML models that use physics-informed, inexpensive descriptors. Their defect ML stage includes a classifier for whether a defect produces a deep in-gap level and a regressor for defect formation energies, which together enable ranking defects by both functionality and synthetic plausibility. Using this pipeline, they report more than 100 promising defect candidates and explicitly shortlist on the order of 100 deep-center defects for quantum-emitter applications (two-level deep centers) and $\sim10$ optimal engineered dopant defects for resistive switching. These approaches can also be applied in Si as a potential future work.

\section{Single emitters in nanophotonic structures} \label{sec:Emitters_in_nanostructures}

After summarizing the properties of single-photon emitters in bulk silicon, we now turn to their integration into nanophotonic structures to implement a spin-photon interface or an ideal single-photon source for quantum applications. Such a device should allow the on-demand creation of a single excitation in a perfectly defined optical field mode. To achieve this with single emitters in silicon, one has to ensure that the emitted light is efficiently collected. In addition, one has to deal with the imperfections of the photon emitters summarized in Sec.~\ref{sec:silicon_emitters}, in particular non-radiative and undesired radiative decay channels (such as PSB for color centers and decay into higher CF levels for Er), as well as fluctuations of the transition frequency owing to the coupling to the environment.

These limitations can be overcome or at least alleviated by confining the light in nanophotonic devices~\cite{gonzalez-tudela_lightmatter_2024}. In this approach, silicon patterning enables the use of interference effects to engineer the LDOS~\cite{lodahl_interfacing_2015} at the emitter position. This can suppress the unwanted transitions and direct the emission to the desired optical mode. In addition, following Fermi's golden rule, an enhanced LDOS leads to faster decay, enabling lifetime-limited photon coherence even for emitters with significant dephasing. 

In this section, we summarize recent results from devices that embed single-photon emitters in nanophotonic silicon structures. We will further discuss the open challenges and possible solutions towards integrated spin-photon interfaces in silicon.

\subsection{Nanophotonic silicon devices for quantum applications} \label{Subsec:NanophotonicDevices}
Over the last few decades, several materials have been explored for the fabrication of nanophotonic devices at the wafer scale, including silicon, silicon carbide, and lithium niobate. Among these, the manufacturing process for silicon is most advanced, owing to its widespread use in semiconductor electronics. However, for photonic applications, the material presents several challenges. First, owing to the small bandgap, low-loss photonics cannot be achieved at visible wavelengths, but requires operation above \SI{1.2}{\micro\meter}. Even there, two-photon absorption can be significant~\cite{sinclair_temperature_2019}, limiting the optical power that can be applied. Second, the large refractive index of silicon, around 3.45 in the telecom bands, results in a high index contrast with both vacuum and silicon dioxide or nitride claddings. As the scattering by surface roughness scales with the index contrast~\cite{melati_real_2014}, propagation losses in silicon waveguides are often larger than in other platforms. In addition, the implementation of photonic crystal devices, which will be discussed in more detail in Sec.~\ref{Subsec:NanophotonicDevices}, requires feature sizes below \SI{100}{\nano\meter} in silicon, which is accessible with electron beam lithography but on the edge of what is feasible with today's small to mid-scale photonic wafer production facilities relying on optical lithography~\cite{panuski_full_2022}. A third challenge of silicon is the realization of active components such as switches and modulators. While heating and free-carrier absorption can be used in classical devices, they cannot be applied in cryogenic quantum devices~\cite{alexander_manufacturable_2025}. Recent approaches to overcome this obstacle include the DC Kerr effect~\cite{chakraborty_cryogenic_2020}, heterogeneous integration with other materials~\cite{eltes_integrated_2020, dong_high-speed_2022}, or microelectromechanics~\cite{errando-herranz_mems_2020, gyger_reconfigurable_2021}.

Despite these challenges, silicon is a widely used material in photonics. The fabrication of corresponding devices typically starts with silicon-on-insulator (SOI) wafers, which are commercially available from several suppliers. The typical device layer (DL) thicknesses are either \SI{0.22}{\micro\meter} ("thin"-SOI) or \SI{2}{\micro\meter} ("thick"-SOI), and the buried oxide (BOX) below the DL typically ranges from 1 to \SI{3}{\micro\meter}. The DL is an ideal starting point for nanofabrication, which typically involves an optical or electron-beam lithography step, followed by reactive-ion etching using bromine, fluorine, or chlorine chemistry. Both thin and thick SOI devices can be obtained from commercial foundries in multi-project wafers. Available processes include spatially selective doping for the implementation of active devices. The DL of foundry-made devices typically resides on the BOX. Alternatively, free-standing structures can be manufactured by selective underetching of the BOX layer, typically in liquid or gaseous hydrofluoric acid (HF), in some cases followed by critical-point drying (CPD) to prevent device failure due to capillary forces. Some of these techniques have been used in the experimental studies described below; the detailed fabrication recipes vary between the experiments and can be found in the corresponding references.

\begin{figure}[tb] 
\includegraphics[width=1.0\columnwidth]{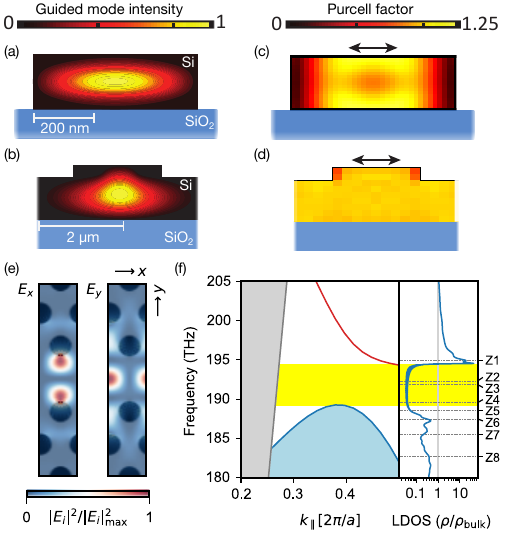}
\caption{   \label{fig:waveguides}
\textbf{Nanophotonic waveguides used for studying photon emitters in silicon}. Color maps of the photonic density of states and electric field profile of light guided in rib, ridge, and photonic-crystal waveguides. The field of the quasi-TE mode in a ridge waveguide (a) is tightly confined, leading to significant variation of the LDOS within the cross-section of the waveguide compared to a bulk crystal. (c) The LDOS is, in turn, directly proportional to the Purcell factor, so that the lifetime of an embedded emitter will depend on its location within the cross-section. In contrast, the larger mode diameter in a single-mode rib waveguide (b) leads to a homogeneous LDOS (d) that is equal to that in bulk. Adapted from Gritsch et al.~\cite{gritsch_narrow_2022}, Physical Review X, 12, 041009, 2022; licensed under a Creative Commons Attribution (CC BY) license. In a photonic crystal (PhC) waveguide, the field is confined to small regions (e). In contrast to the LDOS in rib and ridge waveguides, the LDOS at the field maxima of a PhC waveguide (f) shows a strong spectral dependence. This can suppress undesired optical transitions while simultaneously enhancing the desired one. Adapted from ~\cite{burger_inhibited_2025}.
}
\end{figure} 

\subsubsection{Waveguides}
One of the simplest structures that allow for efficient light extraction from single-photon emitters in silicon is one-dimensional waveguides, which can be efficiently coupled to single-mode fibers using either evanescent, edge- or grating couplers~\cite{dutta_coupling_2016}. We distinguish three types of waveguides:  ridge$\text{-,}$ rib-, and photonic-crystal waveguides. A detailed explanation of these devices can be found in several textbooks, e.g., in Ref.~\cite{vivien_handbook_2013}; therefore, we provide only a brief summary. All three waveguide types have been used with silicon photon emitters, enabling a direct comparison of their properties for emitter integration.

The simplest nanophotonic devices are ridge waveguides, also known as strip waveguides or photonic wire waveguides. Shown in Fig.~\ref{fig:waveguides}a, they are one-dimensional structures that guide light through total internal reflection, typically with a rectangular cross-section of $\SI{0.22}{\micro\meter}$ thickness. This allows single-mode operation below a critical width, such as $\lesssim \SI{0.45}{\micro\meter}$ at 1.55~\textmu m wavelength, featuring two guided modes with orthogonal polarization and quasi-transverse-electrical (quasi-TE, often simplified as TE) and quasi-transverse-magnetic (quasi-TM, TM) character. The modes are non-pure TEM because they have significant field components in the propagation direction. The small dimensions of such waveguides result in strong optical mode confinement, significantly modifying the LDOS, as shown in Fig.~\ref{fig:waveguides}c. At the mode maximum, this enables a high $\beta$-factor, which quantifies the coupling efficiency of embedded emitters into the guided mode. Due to silicon's high refractive index, typical devices achieve $\beta \lesssim 0.8$. In addition, the modification of the LDOS can lead to a measurable slowdown in the decay of a significant fraction of the embedded emitters in the absence of non-radiative relaxation corresponding to Purcell factor of less than one~\cite{gritsch_narrow_2022}. Ridge waveguides have been used for the study of color centers ensembles~\cite{buckley_all-silicon_2017, deabreu_waveguide-integrated_2023}, erbium dopant ensembles~\cite{weiss_erbium_2021, gritsch_narrow_2022, rinner_erbium_2023}, and single G~\cite{prabhu_individually_2023, komza_indistinguishable_2024, buzzi_spectral_2025} and T centers~\cite{lee_high-efficiency_2023, song_entanglement_2025}.

Fabricating ridge waveguides is straightforward, as one only needs to fully etch the masked DL down to the BOX. Their propagation loss is typically in the $\SI{1}{\deci\bel / \centi\meter}$ range~\cite{chrostowski_silicon_2015} and is dominated by scattering caused by surface roughness; record values of $\SI{0.1}{\deci\bel / \centi\meter}$ require special post-fabrication smoothing~\cite{wilmart_ultra_2019}, which also changes the device geometry.  Compared to the other types, ridge waveguides exhibit much smaller bend losses, even for tight turns with a radius of only a few micrometers. This makes them suitable for dense integration and for implementing ring- or racetrack resonators with quality factors exceeding $10^5$.

The second type of device, known as rib waveguides and shown in Fig.~\ref{fig:waveguides}b, is fabricated through controlled partial etching of the DL. It can be used with thin- and thick-SOI wafers; in the latter, typical dimensions are \SI{2}{\micro\meter} in both directions perpendicular to the propagation direction. Appropriately choosing the etch depth allows single-mode operation with two guided modes of perpendicular polarization. The LDOS in such thick-SOI waveguides is not changed compared to bulk crystals in the relevant spectral range~\cite{gritsch_narrow_2022}, see Fig.~\ref{fig:waveguides}d. Thus, the waveguide coupling of embedded emitters only reaches $\beta \lesssim 0.3$. However, because of the small overlap of the guided modes with the etched portion, scattering losses are orders of magnitude smaller than in ridge waveguides and negligible on a chip scale in comparison with bend losses; this enables ring resonators with quality factors exceeding $2\cdot10^7$ in pure silicon devices~\cite{biberman_ultralow-loss_2012}. Low bend loss at tight turns can be achieved in hybrid rib-ridge waveguides, reducing the footprint of corresponding devices~\cite{zhang_compact_2020}.

The third type of waveguides can be engineered by one-dimensional defect modes in photonic crystals. The guided modes in these devices exhibit a pronounced spatial variation of the electric field distribution, as shown in Fig.~\ref{fig:waveguides}e. Typical devices use thin SOI, fully etched to the BOX and then undercut to leave a suspended two-dimensional slab~\cite{baba_slow_2008}. A periodic arrangement of holes in this slab defines the photonic crystal. The used hole pattern determines the dispersion of the guided modes; this allows for a versatile tuning of the local density of photonic states~\cite{lodahl_interfacing_2015, javadi_numerical_2018}, as shown in Fig.~\ref{fig:waveguides}f. This enables a selective suppression of undesired radiative transitions in silicon~\cite{burger_inhibited_2025}, which can already improve the photon-source efficiency for emitters with a high $\eta_{QE}$. In addition, the LDOS at the frequency of the desired transition can be enhanced by the slow-light effect~\cite{baba_slow_2008}. This leads to an expected Purcell enhancement for a perfectly aligned dipolar two-level emitter at the field maximum of~\cite{lodahl_interfacing_2015}:

\begin{equation}
F_{\text{P},\text{TL}} (\omega)= \frac{3 \lambda^2 a}{4\pi n^3 V_\text{eff}}n_g(\omega)
\end{equation}

Here, $n$ is the refractive index, $n_g(\omega)$ is the group index of the mode that quantifies the slowdown of the guided light, and $V_\text{eff}$ is the effective mode volume per unit cell of length $a$. For a $W1$ waveguide, formed by a single, missing row of air holes in a photonic crystal with a hexagonal 2D lattice as shown in Fig.~\ref{fig:waveguides}, this effective mode volume is given as $V_\text{eff}/a=\frac{\lambda^2}{3n^2}$. With previously achieved values~\cite{baba_slow_2008} of $n_g\lesssim 300$ in silicon, this is turn leads to $F_{\text{P},\text{TL}}\lesssim 60$. For emitters with a finite branching ratio and nonradiative decays, this needs to be multiplied by $R_g/R$, i.e., the fraction of the excited state decay in bulk that originates from the desired radiative transitions. An additional reduction factor $\xi<1$ is needed to account for finite polarization and spatial overlap. Still, $F_\text{P}=\xi R_g/R~\cdot F_{\text{P},\text{TL}}$ can significantly exceed one, which would substantially reduce the optical lifetime and enable efficient light extraction from single emitters. However, compared to the other discussed types, slow-light waveguides require more sophisticated design and fabrication procedures and typically exhibit higher losses due to mode matching and fabrication imperfections, particularly when aiming for very high LDOS values~\cite{lodahl_interfacing_2015}.

\subsubsection{Optical resonators} \label{subsubsec:OpticalResonators}

The broadband nature of the above-mentioned waveguide devices makes them well-suited for initial spectroscopy experiments and single-photon generation. However, in the presence of spectral instability in the emitters or significant non-radiative decay, further increasing the LDOS to reduce the radiative lifetime is highly desirable to enable $C>1$. This can be achieved by integrating the emitters into optical resonators via the Purcell effect~\cite{purcell_spontaneous_1946}. At a given wavelength $\lambda$ and refractive index $n$, the Purcell enhancement factor $F_\text{P}$ is determined by the volume $V$ of the resonant mode and the resonator quality factor $Q$:

\begin{equation} \label{eq:Two-Level_PurcellFactor}
F_{\text{P},\text{TL}} = \frac{3}{4\pi^2} \left(\frac{\lambda}{n}\right)^3 \frac{Q}{V}
\end{equation}

This formula assumes a two-level emitter located at the field maximum, with its dipole moment perfectly aligned with the mode's electric field; it thus corresponds to an upper bound. In an actual device, again, this number needs to be reduced by $R_g/R$, and by a factor $\xi<1$ that determines the overlap of the cavity mode with the dipole of the emitter:

\begin{equation} \label{eq:PurcellFactor}
F_\text{P}= F_{\text{P},\text{TL}} \cdot \xi\cdot R_g/R.
\end{equation}

To maximize $F_\text{P}$ and thus $C$, the resonator properties must be optimized within the constraints imposed by the available nanofabrication techniques. To this end, one may attempt to maximize $Q$; however, this comes at a price, as it reduces the device bandwidth. Furthermore, high $Q$ factors require accurate spectral tuning to ensure resonance with the emitter. To mitigate these difficulties, one may rather minimize $V$; however, the smaller the mode volume, the more difficult it becomes to position a single emitter at the mode maximum. In addition, including several spectrally multiplexed emitters in the same silicon resonator~\cite{gritsch_purcell_2023} is hindered at small $V$, as their interactions, which spoil coherence, increase at smaller separations. Therefore, in practice, all devices will face a trade-off between maximizing $F_\text{P}$ and enabling robust manufacturing and operation.

Thus, the first question in the device design is which $F_\text{P}$ is actually required for the coherent emission of a given emitter, i.e., for a cooperativity $C>1$ (see Eq.~\ref{eq:Cooperativity_vs_Purcell}). Based on this, one may determine the best-suited resonator geometry. To this end, different types of resonators can be considered. The first type is based on the rib- or ridge waveguide structures discussed above. By simply forming a closed loop, typical devices achieve $F_\text{P}$ on the order of ten for small ridge waveguide rings~\cite{lefaucher_cavity-enhanced_2023}, and less for high-$Q$ devices with larger mode volumes~\cite{biberman_ultralow-loss_2012, cherchi_dramatic_2013}. With the typically observed dephasing of single-photon emitters in nanofabricated silicon, this means $C \ll 1$.

Thus, higher values of $F_\text{P}$ are desirable. Those can be achieved in photonic crystal cavities, which can achieve mode volumes on the order of the diffraction limit around $(\lambda/n)^3$ with wavelength-scale unit-cell geometries~\cite{dharanipathy_high-q_2014,asano_photonic_2018}. Further reduction can be achieved using tip effects on sub-wavelength scale structures~\cite{hu_design_2016, choi_self-similar_2017}, reaching mode volumes on the order of $10^{-3}(\lambda/n)^3$ at $Q\approx 10^5$ for bow-tie shaped holes in pure silicon without emitters~\cite{hu_experimental_2018}.

To avoid a too-close proximity to interfaces and the difficulty of emitter positioning on such small length scales, instead of minimizing $V$, one may also maximize $Q$ in photonic crystal resonators with larger mode volumes. Record devices have achieved $Q>10^7$ at $V\sim{1.4 \left(\frac{\lambda}{n}\right)^3}$ in pure silicon~\cite{asano_photonic_2017}. Remarkably, also photonic crystal cavities with million-scale $Q$ values can be fabricated using wafer-scale processes~\cite{ashida_ultrahigh-q_2017, panuski_full_2022}, paving the way for the simultaneous operation of many devices on independent chips, which will be further discussed in~\ref{sec:practical_scalability}. Due to fabrication imperfections, this requires post-fabrication tuning, which can be achieved by laser-driven oxidation~\cite{panuski_full_2022} or by gas condensation at cryogenic temperatures, followed by laser-controlled melting~\cite{zeng_cryogenic_2023, gritsch_purcell_2023}.

\begin{figure}[tb] 
\includegraphics[width=1.0\columnwidth]{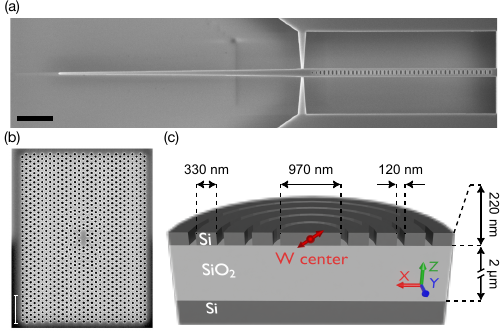}
\caption{     \label{fig:cavities}
\textbf{Emitters in nanophotonic cavities} (a)  Scanning probe microscopy (SEM) picture of a free-standing photonic crystal cavity with embedded erbium dopants. The length of the scale bar is \SI{3}{\micro\meter}. Adapted from Gritsch et al.~\cite{gritsch_optical_2025}, Nature Communications, 16, 64, 2025; licensed under a Creative Commons Attribution (CC BY) license. (b) SEM picture of a two-dimensional photonic-crystal cavity incorporating a single G center. The length of the scale bar is \SI{2}{\micro\meter}. Adapted from Saggio et al.~\cite{saggio_cavity-enhanced_2024}, Nature Communications, 15, 5296, 2024; licensed under a Creative Commons Attribution (CC BY) license. (c) Schematic of circular-Bragg-grating cavity including a single W center in its field maximum. Adapted from Lefaucher et al.~\cite{lefaucher_bright_2025},  	arXiv:2501.12744, 2025; licensed under a Creative Commons Attribution (CC BY) license. }
\end{figure}

To use such devices as single-photon sources, one needs to ensure efficient in- and out-coupling of the emission. Fig.~\ref{fig:cavities} shows the designs used in recent demonstrations. In Fig.~\ref{fig:cavities}a, a nanobeam photonic crystal cavity is defined using a perturbed line of elliptical holes~\cite{gritsch_optical_2025}. A section of three holes, adiabatically reduced in size, couples the photonic crystal cavity to the feed waveguide, which ends at a tapered coupler for off-chip coupling. The cavity features imbalanced reflectivities to increase the outcoupling towards the coupler side. Similar waveguide-coupled designs have been used with single erbium dopants~\cite{gritsch_purcell_2023, gritsch_optical_2025} and color centers~\cite{islam_cavity-enhanced_2024, johnston_cavity-coupled_2024, komza_multiplexed_2025, photonic_inc_distributed_2024}.

Instead of on-chip coupling, one can design the resonator such that it preferentially emits into a free-space mode that can be collected efficiently~\cite{minkov_photonic_2017, asano_photonic_2018}. Examples with embedded silicon emitters are the two-dimensional photonic-crystal cavities~\cite{redjem_all-silicon_2023, saggio_cavity-enhanced_2024} shown in an SEM image in Fig.~\ref{fig:cavities}b~\cite{saggio_cavity-enhanced_2024}, or the circular grating cavities~\cite{lefaucher_bright_2025} schematically depicted in Fig.~\ref{fig:cavities}c. The latter structure was fabricated on top of a previously selected individual emitter, ensuring the emitter was optimally positioned within the cavity. This technique enabled emitter-cavity integration at a comparably low emitter concentration, thereby minimizing interactions and damage to the DL that may result from the emitter integration process. However, its implementation requires emitters that are bright enough to be detected in a plane layer; while this works for bright color centers, it may be difficult with erbium dopants.

In summary, numerous approaches towards the Purcell-enhanced operation of single emitters in silicon have been explored. The next sections will summarize the properties of individual emitters found in corresponding devices.

\subsection{Single emitter properties in nanophotonic silicon devices}
\label{sec:optical}

The performance of photon emitters is determined by the efficiency of generating single photons that are indistinguishable in all their degrees of freedom \cite{esmann_solid-state_2024}. In this context, nano-patterning can be a double-edged sword. On the one hand, it allows tailoring the optical LDOS in nanostructures, thereby enhancing light extraction efficiency and reducing the lifetime, thus relaxing the requirements on optical coherence for Fourier-limited emission. On the other hand, the proximity of interfaces and defects introduced during nanofabrication can also compromise the spectral stability of the emitters. Therefore, experimental investigations are required to evaluate the performance of single-photon emitters and spin-photon interfaces in silicon. The following sections will describe recent results in this context.

\subsubsection{Lifetime reduction and photon extraction}  \label{sec:optical:optical}

After describing the different types of nanophotonic silicon resonators in Sec.~\ref {subsubsec:OpticalResonators}, we will summarize the lifetime reductions achieved in recent experiments. An overview is provided in Table~\ref{tab:emittersincavity}, along with the resonator type, mode volume, quality factor, and observed Purcell enhancement (determined from the lifetime reduction). The table focuses on experiments that integrated single emitters in silicon nanophotonic cavities; it does not include recent ensemble measurements in microdisks, Bragg gratings, and ring resonators~\cite{lefaucher_cavity-enhanced_2023, lefaucher_purcell_2024, chiles_purcell-enhanced_2024, veetil_enhanced_2025}, as well as Mie resonators~\cite{ahamad_controlling_2025} and sub-micron islands~\cite{khoury_light_2022, higginbottom_optical_2022}.

In silicon photonic crystal devices without embedded emitters, $Q>10^7$ has been achieved for mode volumes of $\sim (\lambda/n)^3$~\cite{asano_photonic_2018}, which would lead to a potential maximum Purcell enhancement approaching $F_{\text{P},\text{TL}}\approx10^6$. However, integrating color centers or erbium dopants by implantation can significantly reduce $Q$. While post-implantation annealing procedures can recover the crystalline structure, the introduced impurities and defects may remain a significant source of loss through optical absorption. Thus, the first demonstrations with color centers and erbium dopants only reached $Q\lesssim10^5$~\cite{redjem_all-silicon_2023, gritsch_purcell_2023, islam_cavity-enhanced_2024,johnston_cavity-coupled_2024, saggio_cavity-enhanced_2024, photonic_inc_distributed_2024, komza_multiplexed_2025}. Only recently has $Q>10^6$ been achieved with Er:Si; however, the experiment used a lower dopant concentration in a Fabry-Perot cavity rather than a nanophotonic resonator. The larger mode volume resulted in a lower $F_\text{P}<3$~\cite{frueh_spectral_2026}. 

\begin{figure}[tb] 
\includegraphics[width=1.0\columnwidth]{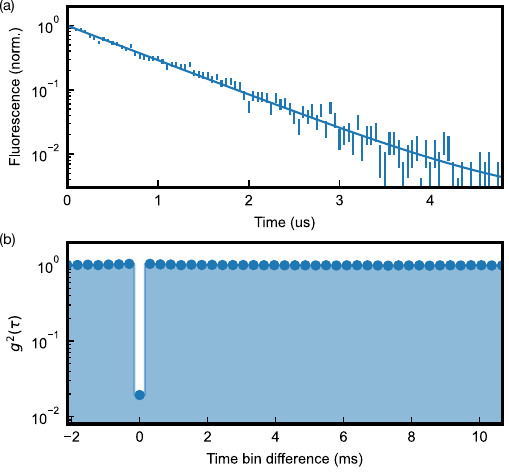}
\caption{     \label{fig:g2_PurcellEnhancedDecay}
\textbf{Purcell-enhanced single photon emission.} The measurements are performed on a single erbium dopant in a one-dimensional photonic crystal cavity with a mode volume of $0.83\,(\lambda/n)^3$ and a quality factor of $82\times10^3$. (a) The Purcell effect reduces the lifetime of the dopant from its bulk value of $\SI{142}{\micro\second}$ to $\SI{0.8}{\micro\second}$ (blue data and exponential fit), corresponding to a $F_\text{P}=177$-fold enhancement of the decay rate. (b) The autocorrelation function $g^{(2)}(\tau)$, measured with a single detector, exhibits clear antibunching and reaches $0.019$ in the time bin corresponding to zero delay $\tau$. Adapted from Gritsch et al.~\cite{gritsch_optical_2025}, Nature Communications, 16, 64, 2025; licensed under a Creative Commons Attribution (CC BY) license.
}
\end{figure}

In comparison, $F_\text{P}=177$ has been demonstrated in a nanophotonic silicon resonator with Er:Si~\cite{gritsch_optical_2025}, as shown in Fig.\ref{fig:g2_PurcellEnhancedDecay}a. This value was limited by the finite $\eta_{BR}=0.23$ and field-emitter overlap $\xi$. With color centers, the finite DWF and quantum efficiencies $\eta_{QE}$ (see Tab.~\ref{tab:colorcenters}) lead to a slightly lower observed $F_\text{P}\lesssim 61$ at comparable resonator parameters~\cite{photonic_inc_distributed_2024, komza_multiplexed_2025, redjem_all-silicon_2023}, see table~\ref{tab:emittersincavity}. Nevertheless, substantial lifetime reductions have been achieved with the T, G, $\text{G}^*$, and W centers.

In the future, the achieved $F_\text{P}$ values may be further increased by optimized materials and fabrication procedures. So far, the smallest $V$ of resonators with embedded emitters~\cite{johnston_cavity-coupled_2024, islam_cavity-enhanced_2024} has reached $\sim 0.2 \left( \frac{\lambda}{n} \right)^3$. This may still be reduced by two orders of magnitude~\cite{hu_experimental_2018}, in case single emitters can be successfully integrated into such devices, and if the resulting proximity to surfaces does not hamper the optical or even the spin coherence of the emitters, e.g., by inducing energy transfer from the emitter to the surface~\cite{liu_quantification_2023}. This can lead to enhanced non-radiative decay in nanostructures due to higher recombination rates~\cite{aberle_surface_2000}, which has been suggested for W centers~\cite{baron_detection_2022} integrated at a depth of $\SI{60}{\nano\meter}$.

\begin{table*}
\begin{ruledtabular}
\begin{tabular}{ccccccccccc}
 Emitter & Cavity & $\Gamma_{\mathrm{bulk}}$ (ns) & $\Gamma_{\mathrm{nano}}$ (ns) & $F_\text{P}$ & Q & V$\left(\lambda/n\right)^3$ & $g^{\left(2\right)}(0)$ & $\sigma_\text{SD} \text{(GHz)}$ & $\eta$ &  Ref.\\
 \hline
 Er (site B) & 1D PhC & \num{186e3} & 2400 &  78 & \num{73e3} & 1.45 & 0.39 & 0.08 & \SI{5}{\percent} & \cite{gritsch_purcell_2023}\\
 Er (site A) & 1D PhC & \num{142e3} &  800 & 177 & \num{82e3} & 0.83 & 0.019 & 0.014& \SI{20}{\percent} & \cite{gritsch_optical_2025}\\
 G center    & 1D PhC & 6 & 0.97 & 31 & \num{4.6e3} & 0.26 & 0.408 & 17 &\SI{37.5}{\percent} & \cite{kim_bright_2025}\\
             & 2D PhC & 6 & 7 & 1 & \num{3e3} & 1 & 0.03 & $\sim 10^{\ast}$ & & \cite{saggio_cavity-enhanced_2024} \\
             & Bragg grating & 10.27 & 4.13 & 11 & \num{0.2e3} & & 0.37 & & & \cite{ma_nanoscale_2025}\\
 G$^{*}$ center   & 2D PhC & 53.6 & 6.7 & 29 & \num{3.2e3} & 0.66 & 0.30 & 6.8 &\SI{30}{\percent} & \cite{redjem_all-silicon_2023}\\
 T center    & 1D PhC & 940 & 62 & 61 & \num{35e3} & 0.5 & 0.0079 & $1.6^{\ast\ast}$ & \SI{32}{\percent} & \cite{komza_multiplexed_2025}\\
             & 1D PhC & 912 & 168.7 & 5.4 & \num{6.6e3} & 0.23 & 0.07 & & \SI{23}{\percent} & \cite{islam_cavity-enhanced_2024}\\
             & 1D PhC & 940 & 64.5 & 14.6 & \num{25.65e3} &  & 0.0076 & 1.1 & & \cite{photonic_inc_distributed_2024}\\
             & 1D PhC & 835.2 & 136.4 & 6.89 & \num{43e3} & 0.172 & 0.024 & 3.8 &\SI{5}{\percent} & \cite{johnston_cavity-coupled_2024} \\
 W center    & Bragg grating & 34 & 7 & 7.2 & \num{0.158e3} & & 0.03 & & \SI{21}{\percent} & \cite{lefaucher_bright_2025}\\
\end{tabular}
\end{ruledtabular}
\caption{Emitters in nanophotonic silicon resonators. The table summarizes recent experiments with different emitters and different resonator geometries (PhC: Photonic crystal), as described in~\ref{subsubsec:OpticalResonators}. The Purcell factor $F_\text{P}$ is calculated by comparing the lifetime in the resonator $\Gamma_\text{nano}$ to that in bulk silicon crystals $\Gamma_\text{bulk}$. The probability of emitting two photons at the same time is quantified by the autocorrelation function at zero time delay $g^{(2)}(0)$. $\sigma_\text{SD}$ is the observed spectral diffusion linewidth (FWHM) of the narrowest emitter, and $\eta$ is the achieved efficiency of generating a photon on demand into a single-mode fiber. $\ast$: Extracted graphically. $\ast\ast$: Power-broadened value.\label{tab:emittersincavity}}
\end{table*}

The lifetime reduction observed across the different silicon emitters will lead to enhanced single-photon emission rates. The demonstrated efficiency for generating a single photon in a fiber-coupled mode on demand can thus exceed $30\,\%$. Further improvements can be achieved by utilizing over-coupled resonators and optimizing waveguide-to-fiber~\cite{zeng_cryogenic_2023} or cavity-to-free-space~\cite{panuski_full_2022} coupling designs.

Besides its efficiency, the multiphoton probability is a key figure of merit of single-photon sources. Table~\ref{tab:emittersincavity} also includes the experimental values, which are determined by measuring the autocorrelation function of the emitters at zero time delay. A typical measurement, done with Er:Si, is shown in Fig.~\ref{fig:g2_PurcellEnhancedDecay}b. The best values achieved so far are on the order of one or a few percent. While this unambiguously demonstrates that the experiments are performed on single emitters, a further reduction will be required for high-fidelity quantum information processing operations. This will likely be achievable by device engineering after a systematic study on the sources of multiphoton emission.

\subsubsection{Spectral stability}
To utilize single photon sources in linear-optics-based photonic and distributed quantum information processing, they must be indistinguishable in all their properties, or at least exhibit a spectrally stable emission frequency, i.e., a Fourier-limited optical coherence, such that $C\gg1$. In single emitters, the coherence of their ground and excited states is directly transferred to that of the photons. As explained in more detail in Sec.~\ref{sec:silicon_emitters}, the coupling to the environment can lead to random shifts of the transition frequency, leading to homogeneous broadening and spectral diffusion of the emitter, depending on the timescales.
 
While very stable emission frequencies with inhomogeneous linewidths below $\SI{50}{\mega\hertz}$ have been observed with color centers in isotopically purified bulk crystals~\cite{chartrand_highly_2018, bergeron_silicon-integrated_2020}, the spectral stability observed in nanophotonic devices is much worse. The spectral diffusion of cavity-integrated color centers is on the scale of a few $\si{\giga\hertz}$~\cite{redjem_all-silicon_2023, photonic_inc_distributed_2024, saggio_cavity-enhanced_2024, johnston_cavity-coupled_2024, islam_cavity-enhanced_2024, komza_multiplexed_2025}. With resonance checks, this can be reduced approximately tenfold~\cite{bowness_laser-induced_2025,zhang_laser-induced_2025}, but still exceeds the lifetime-limit approximately hundredfold. The dephasing observed on short timescales~\cite{zhang_laser-induced_2025, bowness_laser-induced_2025} can be significantly smaller, which places current experiments~\cite{photonic_inc_distributed_2024, komza_multiplexed_2025} around $C\approx 0.1$.

Much narrower SD values of $\sim\SI{20}{\mega\hertz}$ are found in nanophotonic Er:Si~\cite{burger_inhibited_2025, gritsch_optical_2025}, where the filled outer 5s and 5p shells shield the inner 4f electrons from electric fields. The optical coherence of single emitters can even exceed microseconds in nanophotonic devices, and $\SI{20}{\micro\second}$ in $\SI{2}{\micro\meter}$ thick membranes~\cite{frueh_spectral_2026}, exceeding those of single color centers in nanostructures~\cite{zhang_laser-induced_2025, bowness_laser-induced_2025} by more than three orders of magnitude. However, the cavity-enhanced emission is also tenfold slower. Still, current Er:Si photonic crystal cavities~\cite{gritsch_optical_2025,frueh_spectral_2026} achieve $C\approx1$.

In both color center and erbium based devices, the observed broadening of the homogeneous and spectral diffusion linewidths is far too large to be explained by magnetic interactions between the emitters and other impurities or the host's nuclear spins. Instead, it is attributed to electric-field noise caused by fluctuating charges at device interfaces, by weakly bound charges associated with impurities and defects in the material, or by accumulated charges due to electrical structures on the silicon chip~\cite{candido_suppression_2021}. These noise sources can be more pronounced in nanostructures for three reasons: First, in nanoscale devices, the emitters are necessarily in close proximity to interfaces. Second, the fabrication processes used may cause crystal damage and increase impurity concentrations. In particular, oxygen and hydrogen are expected to be present in smart-cut fabricated SOI, and other atomic species used in the nanopatterning process may diffuse into the material. Finally, to generate single emitters in submicron mode volumes, one needs a much higher emitter concentration; as a consequence, the structural and chemical integrity of the crystal is much worse than that of ultrapure bulk material.

Which of these charge noise sources is dominant remains an open question and will strongly depend on the device geometry and fabrication procedure~\cite{frueh_spectral_2026}. However, for all structures, some improvement may be expected by two approaches. First, surface passivation can stabilize the charge environment at the interfaces. It has been used in photovoltaics to reduce surface recombination and Auger effects \cite{aberle_surface_2000,bonilla_dielectric_2017}, in nanophotonic structures to smoothen the edges and reduce scattering loss \cite{borselli_measuring_2006,borselli_surface_2007}, and in photodetectors to suppress dark currents \cite{mayet_surface_2018,gherabli_role_2020}. Second, applying a strong bias field can deplete charge traps at both interfaces and in the bulk, thereby stabilizing the charge environment \cite{candido_suppression_2021, anderson_electrical_2019}. In silicon nanophotonics, controlled application of electric fields can be achieved in co-integrated diode structures formed by localized doping. This allows stabilizing the charge state of single-photon emitters and, potentially, their surroundings, and may be used to tune their emission frequency~\cite{day_electrical_2024}.

Recent measurements on laser-induced dephasing have demonstrated that the resonant pulses used to optically excite the emitters lead to a reconfiguration of the charge environment and thus to instantaneous spectral diffusion~\cite{bowness_laser-induced_2025,zhang_laser-induced_2025, frueh_spectral_2026}. The exact sources and mechanisms need further investigation to improve the spectral stability. In the meantime, the implementation of a resonance-check pulse, as originally used with color centers in diamond~\cite{ruf_quantum_2021}, can be employed to mitigate the detrimental impact of laser-induced spectral diffusion on long timescales. Alternatively, a Fourier-limited SD linewidth may be achieved by increasing the Purcell factor; the required improvement factor is $\sim 10$ in Er:Si, and 100 to 1000 in color-center devices. With this, the strong temporal filtering required to achieve a high visibility in the first experiments on two-photon interference~\cite{komza_indistinguishable_2024, photonic_inc_distributed_2024} may not be required in the future, paving the way for distributed quantum information processing at high rates and high fidelity.

\begin{figure}[tb] 
\includegraphics[width=1.0\columnwidth]{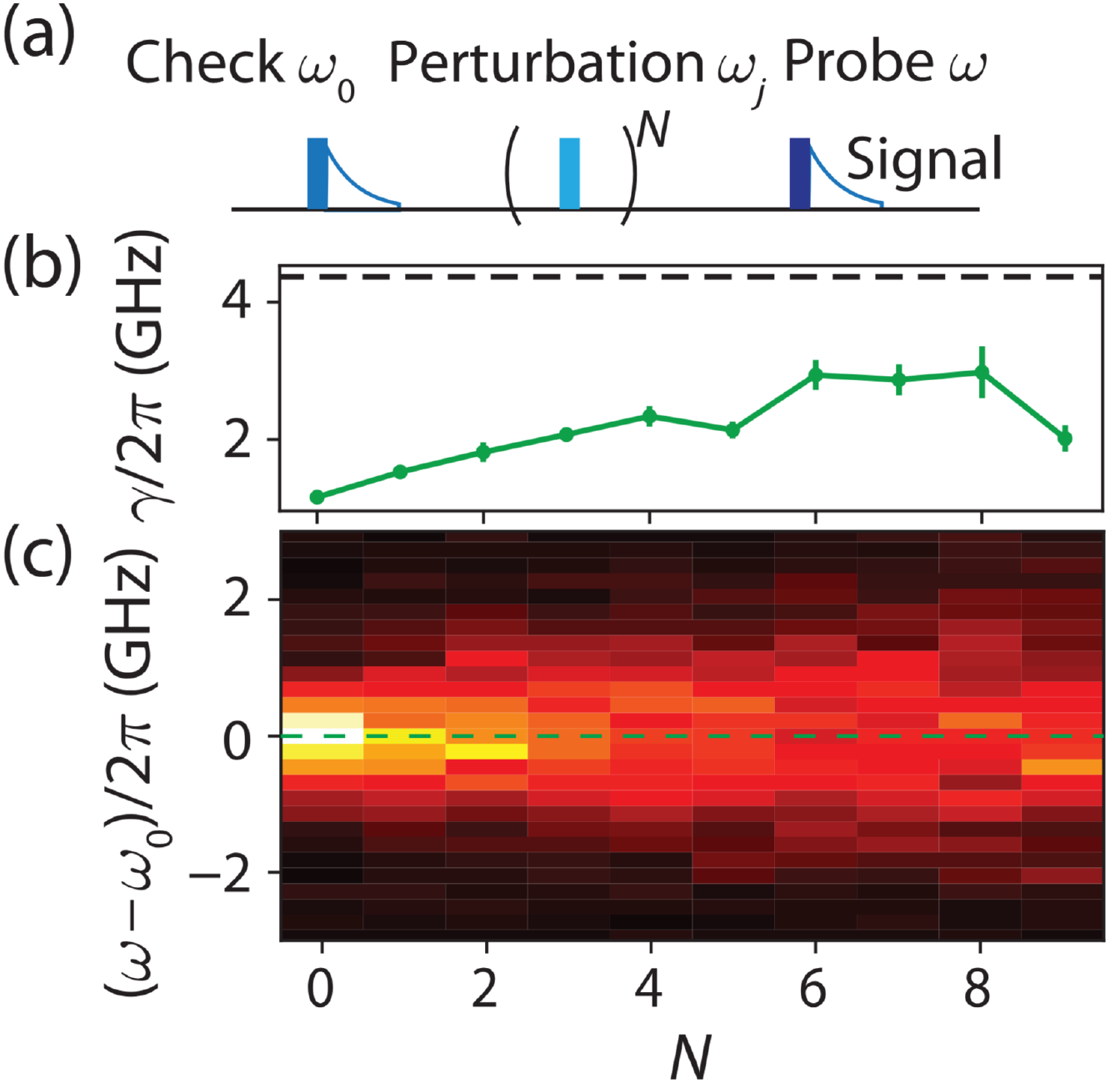}
\caption{     \label{fig:Laser_induced_SD}
\textbf{Laser-induced spectral diffusion.} (a) Check-probe pulse sequence. Detecting an emitted photon from a single T center in a nanophotonic cavity after a first "check" laser pulse allows one to narrow down its emission frequency within the spectral diffusion linewidth. The probability of detecting a photon at the same frequency in a subsequent "probe" pulse will drop in case the emitter frequency has changed. Inserting $N$ perturbation pulses enables the study of their effects on SD. (b) The linewidth $\gamma$ of the emitter increases with $N$, showing that the laser pulses induce instantaneous spectral diffusion. The broadening can also be seen in the line shape (c). Adapted from Zhang et al.~\cite{zhang_laser-induced_2025}, PRX Quantum, 6, 030351, 2025; licensed under a Creative Commons Attribution (CC BY) license.
}
\end{figure}

\subsubsection{Spin properties in nanophotonic spin-photon interfaces}

While the previous sections focused on the optical properties of emitters in nanophotonic silicon devices, in the following, we will briefly summarize the spin properties of nanoscale devices, which are central to distributed quantum information processing and quantum networking. The investigation of spin qubits in silicon has a long history. Early investigations include electron paramagnetic resonance measurements, where, e.g., the electron spins of phosphorus donors in silicon exhibited coherence times on the order of \SI{100}{\micro\second} in bulk samples~\cite{gordon_microwave_1958}. The field has grown rapidly since it was proposed that quantum computers could be implemented using nuclear spin qubits associated with donors in silicon~\cite{kane_siliconbased_1998}. Key results in this context included the electronic high-fidelity readout of single nuclear spins~\cite{pla_high-fidelity_2013} and the observation of hour-long nuclear spin coherence up to room temperature in isotopically purified material~\cite{saeedi_room-temperature_2013}. The results of these and related experiments are summarized in recent review articles, including~\cite{zwanenburg_silicon_2013, morello_donor_2020, burkard_semiconductor_2023}. 

These previous experiments all used electronic circuits for the control and readout of single spins in silicon. In contrast, using optical control fields offers several advantages. First, optical frequencies offer superior bandwidth, enabling frequency-multiplexed addressing of hundreds of spin qubits within a few cubic micrometers~\cite{ulanowski_spectral_2024}. Second, spin-photon interfaces allow the distribution of quantum states over long distances using optical fibers. Third, optical fields can be confined to the nanoscale, eliminating cross-talk even between closely-spaced photonic components. Finally, fast and efficient photon detectors can be integrated on silicon chips~\cite{akhlaghi_waveguide_2015} to enable rapid, low-noise spin qubit measurements via spin-to-optical conversion~\cite{chatterjee_semiconductor_2021}.

To this end, the spins of both single T centers~\cite{higginbottom_optical_2022} and single erbium dopants~\cite{gritsch_purcell_2023} have been investigated. Integrating these emitters into nanophotonic waveguides and resonators has improved the efficiency of photon outcoupling, which has facilitated the optical readout of the electronic spin of single erbium dopants~\cite{gritsch_optical_2025} and the nuclear spin of single T centers~\cite{photonic_inc_distributed_2024, song_entanglement_2025}. Reported state preparation and measurement (SPAM) fidelities reach $0.869(8)$~\cite{gritsch_optical_2025} for Er:Si and $0.89(6)$ for T centers~\cite{photonic_inc_distributed_2024}. The achieved numbers are limited by electron-spin flips upon optical excitation~\cite{gritsch_optical_2025, bowness_laser-induced_2025}, and may be improved by devices with higher Purcell enhancement, increased outcoupling, and higher photon-detection efficiencies. This will be crucial to cross the fault-tolerance threshold and thus enable up-scaling.

Towards this end, long coherence times and high-fidelity spin rotations are also required. The latter can be achieved by irradiating microwave fields that drive the spin transitions, i.e., using the same techniques as in electronically controlled spin qubits. All-optical qubit control via detuned high-power laser fields in a lambda-configuration to drive Raman transitions~\cite{goldman_optical_2020} may be considered as an alternative that requires lower drive powers and eliminates cross-talk. However, this may be hampered in silicon by the laser-induced dephasing described earlier~\cite{bowness_laser-induced_2025,zhang_laser-induced_2025, frueh_spectral_2026}.

The spin coherence of the mentioned photon emitters in silicon has been investigated using Ramsey and spin-echo experiments, which allow determining the dephasing time $T_2^*$ and the echo time $T_\text{Hahn}$, respectively. In isotopically purified material, ensembles of both T centers in bulk crystals~\cite{bergeron_silicon-integrated_2020} and erbium dopant ensembles in nanoscale devices~\cite{berkman_long_2025} exhibited $T_\text{Hahn}\gtrsim \SI{1}{\milli\second}$ for the electron spin at small magnetic bias fields of $\SI{10}{\milli\tesla}$ and cryogenic temperatures $<\SI{0.3}{\kelvin}$. 

The integration of these emitters into nanostructured devices may increase the magnetic field noise. As an example, experiments with electrically-controlled spins found paramagnetic impurities at interfaces \cite{de_sousa_dangling-bond_2007, paik_t_1_2010, rosskopf_investigation_2014}. Similarly, implantation-induced crystal damage may lead to the formation of paramagnetic impurities. In effect, in first single-emitter experiments~\cite{photonic_inc_distributed_2024} in a photonic crystal cavity using isotopically purified SOI, the coherence of T center electronic spins decreased by roughly an order of magnitude to $T_\text{Hahn}=\SI{0.27}{\milli\second}$ at $\SI{1.6}{\kelvin}$ and $\SI{0.1}{\tesla}$. In silicon with natural isotopic abundance, $T_\text{Hahn}=\SI{0.05}{\milli\second}$ for the electronic spin of single erbium dopants~\cite{gritsch_optical_2025} and $\SI{0.4}{\milli\second}$ with single T centers. These values are close to the expectation~\cite{witzel_quantum_2012} based on the coupling to the slowly-fluctuating bath of $^{29}\text{Si}$ nuclear spins. 

Interactions with nuclear spins and paramagnetic impurities can be dynamically decoupled by suited microwave pulse sequences~\cite{suter_colloquium_2016}. With this, a strong increase in the spin coherence time $T_2 \gg T_\text{Hahn}$ can be expected, up to the limit imposed by the spin lifetime $T_2 < 2 T_1$ in case the pulses exhibit a high fidelity. First experiments with T centers achieved $T_2\gtrsim \SI{1}{\milli\second}$ for the electron spin, limited by the pulse fidelity~\cite{song_entanglement_2025}.

In samples with a concentration of resonant spins, dynamical decoupling requires tailored sequences to simultaneously eliminate the detrimental effects of disorder, magnetic-field noise, and spin-spin interactions ~\cite{zhou_robust_2023}. However, perfect decoupling of the latter is not possible for spins with anisotropic g-tensors~\cite{merkel_dynamical_2021}, including Er:Si~\cite{holzapfel_characterization_2025}, which may set a limit on the usable emitter concentrations. 

In low-concentration samples, the spin lifetime will not be limited by spin-spin interactions, but by spin-lattice relaxation. Depending on the temperature and magnetic field, phononic relaxation via the direct, Raman- or Orbach-processes can dominate~\cite{wolfowicz_quantum_2021}. Lifetimes exceeding tens of seconds have been achieved at small fields and ultralow temperatures in nanoscale silicon devices~\cite{berkman_long_2025}. In contrast, experiments on T centers in nanophotonic devices~\cite{higginbottom_optical_2022,deabreu_waveguide-integrated_2023} have shown decreased lifetimes compared to bulk values of $>\SI{1}{\second}$~\cite{bergeron_silicon-integrated_2020}, which was attributed to laser-induced spin flips.

Spin-lattice relaxation is determined by the phonon density of states. In silicon nanobeams, this has been measured and calculated for thermal transports~\cite{park_phonon_2017, yang_mean_2013}; based on this, no lifetime improvements are expected in common devices. However, longer lifetimes can be expected in nanostructures with a tailored DOS that exhibits a bandgap at the spin transition frequencies. This effect has recently been demonstrated with color centers in diamond~\cite{kuruma_controlling_2025}; adapting it to silicon devices may pave the way for coherent operation at significantly elevated temperatures. 

Nuclear spins of the defects or proximal $^{29}\text{Si}$ atoms may be used as qubit registers. Recent experiments~\cite{song_entanglement_2025} have demonstrated the entanglement of the H and C spins of a T center in a nanophotonic waveguide with $77\,\%$ fidelity, using techniques pioneered in earlier experiments with color centers in diamond~\cite{dutt_quantum_2007, robledo_highfidelity_2011}. In such a nuclear spin register, one may store quantum information for longer times because of the negligible coupling of nuclear spins to phonons and their weaker magnetic moments, which are approximately four orders of magnitude smaller than those of electronic spins. With nuclear spins of donors in silicon, coherence times exceeding \SI{44}{\second} have been measured at \SI{1.7}{\kelvin} \cite{steger_quantum_2012}. The hydrogen nuclear spins associated with cavity-coupled T centers in isotopically purified samples, however, only reached \SI{0.22}{\second} in the first experiments~\cite{photonic_inc_distributed_2024}. Nevertheless, such coherence times are already promising for the implementation of distributed quantum information processing. A first experiment in this direction was the entanglement of individually addressed T centers in nanophotonic cavities, although with limited entanglement fidelity (0.6) and rate (20~mHz) arising from broadened homogeneous linewidths and limited cavity enhancement~\cite{photonic_inc_distributed_2024}. These values will need to be improved for distributed quantum information processing. For this, it will be required that the nuclear spin coherence is preserved upon optical excitation. This may be achieved using T centers, as concluded from a recent investigation of their hyperfine interaction~\cite{brunelle_silicon_2025}. 

In summary, spin control has been achieved with both the T center and erbium dopants. Future experiments should aim at improving the state-preparation and measurement fidelity. In addition, improving the optical coherence of the corresponding spin-photon interfaces is paramount to enabling high-rate, high-fidelity entanglement between remote single-photon emitters.

\section{Practical scalability}   \label{sec:practical_scalability}
The previous sections have summarized the physical properties of single emitters and spin-photon interfaces in silicon. In the following, we describe their potential for upscaling and the potential limitations that need to be overcome to this end.

\subsection{Application-driven requirements} ~\label{subsec:Applications}
Solid-state emitters underpin a wide range of quantum technologies, including sensing, communication, networking, and information processing. While these applications share common hardware elements -- optical transitions, spin degrees of freedom, and photonic interfaces -- they impose fundamentally different system-level requirements. Understanding these requirements is essential for evaluating which performance metrics are critical for scalability, independent of the specific emitter and experimental implementation.

\subsubsection{Quantum sensing}

Quantum sensing describes the use of a quantum system or quantum phenomena to perform measurements of a physical quantity~\cite{degen_quantum_2017, pirandola_advances_2018}. It has reached a high level of maturity in photonic, atomic, and solid-state platforms, with applications ranging from biophysics~\cite{aslam_quantum_2023} to fundamental physics research, including the exploration of novel constituents of matter~\cite{ye_essay_2024}. Solid-state quantum sensors are typically based on the spin of color centers such as nitrogen-vacancy centers in diamond~\cite{barry_sensitivity_2020} or emitters in van-der-Waals materials~\cite{fang_quantum_2024}, where single-emitter sensitivity and robustness at room temperature are well established. In these and related systems, the key challenge for quantum sensing is not single-emitter performance but scalability: increasing spatial coverage, parallelism, and integration while maintaining sensitivity.

While many different solid-state platforms are being explored, silicon-based systems may offer a complementary pathway toward scalable sensing architectures. Dense arrays of emitters integrated with photonic routing and low-loss microwave structures would enable parallel, spatially resolved measurements and on-chip signal processing. Clearly, optical coherence or deterministic photon-mediated interactions may be less critical in such applications. Instead, the key requirements are reproducible emitter formation, stable spin control, and scalable readout. In this context, a key limitation of the emitters discussed in this article is that long spin coherence times require cryogenic conditions. Instead, operating at elevated temperatures would be favorable for practical or biological applications~\cite{aslam_quantum_2023}. With the currently known emitters, this is only possible in case optical properties are used instead of spin states. A first example based on Er:Si is temperature measurements in silicon nanostructures up to ambient temperature~\cite{sandholzer_luminescence_2025}. However, this relied on fluorescence changes rather than quantum coherence.

However, the requirement of room-temperature operation can be --- and likely has to be -- dropped when targeting the second key open challenge of quantum sensing: The use of large entangled states to push sensor sensitivity to its ultimate limits. Here, the vision is to leverage advances in quantum information science to define and advance the frontiers of measurement physics, opening new possibilities for both applied and fundamental physics~\cite{ye_essay_2024}. Implementing such systems, including distributed measurement approaches, puts stringent demands on the qubits coherence. The detailed requirements will depend on the specific application and protocol, but will be very similar to those encountered in quantum information processing, which will be discussed in the following sections.

\subsubsection{Quantum communication and networking}

A major goal of quantum communications is the distribution of a secret key between remote parties~\cite{gisin_quantum_2007}. However, there are also many other applications that become feasible once entangled states can be shared between distant users in a quantum internet~\cite{wehner_quantum_2018}. To achieve practical rates in such a system, the two key requirements are low optical loss and massive multiplexing, enabled via photonic integration. 

Owing to their operation in the telecommunications frequency bands (see Fig.~\ref{fig:overview}), silicon emitters are perfectly suited to achieve low loss over both fiber- and free-space channels~\cite{holewa_solid-state_2025} and are compatible with existing classical communication infrastructure. However, the residual absorption in optical fibers will still limit the distance over which direct communication is feasible to a few hundred kilometers. This limitation can be overcome by quantum repeaters~\cite{briegel_quantum_1998}. Corresponding first- and second-generation devices~\cite{muralidharan_optimal_2016} require long-term storage and processing capabilities, which are readily available in silicon spin-photon interfaces that allow the combination of stationary memory and flying communication qubits. 

Because of the mentioned losses, entanglement generation over long distances is inherently probabilistic. Thus, scalability relies heavily on parallelism rather than determinism, and spectral, spatial, and temporal multiplexing of many emitters is required to overcome low success probabilities and achieve practical entanglement distribution rates. In this context, it may be noted that compared to quantum dots~\cite{holewa_solid-state_2025}, the typical timescales for photon emission in silicon devices are significantly longer. However, with frequency rather than time-domain multiplexing, this will not affect the achievable communication rates, which are typically limited by the signaling time~\cite{muralidharan_optimal_2016}, i.e., the time required to transmit photons rather than generate them. First- and second-generation quantum repeater architectures will exhibit nodes at least tens of kilometers apart --- at a distance where photon loss dominates system inefficiency and imperfections. This corresponds to signaling times exceeding $\SI{0.1}{\milli\second}$. All devices discussed in this review emit photons on much shorter timescales. Thus, a very strong Purcell enhancement may not be required, and unnecessarily raise the complexity of qubit control.

Instead, key system-level performance metrics include spectral stability over the timescale of photon emission, spin--photon entanglement fidelity, memory coherence during repeated remote entanglement attempts, and efficient coupling to photonic components. Importantly, moderate efficiencies of the spin-photon interfaces can be sufficient when combined with massive multiplexing, as unsuccessful attempts can be repeated without introducing logical errors. Furthermore, devices targeted at secret-key distribution or remote entanglement can tolerate significant error rates in remote operations when high-fidelity local ones are available for entanglement distillation~\cite{muralidharan_optimal_2016}. Still, achieving practical rates will strongly benefit from high-efficiency, high-fidelity remote entanglement~\cite {loock_extending_2020}.

\subsubsection{Distributed quantum computing}

Distributed quantum computing architectures are closely related to the goals and techniques of quantum networks described above. The key idea is to connect separated quantum processors by photonic links~\cite{simmons_scalable_2024}. Compared to monolithic architectures, the increased connectivity can improve the system resilience and enable error-correction protocols with significantly relaxed requirements~\cite{breuckmann_quantum_2021}. Again, the emitters in such systems play a dual role as sources of single photons and as optically interfaced quantum memories that store and process quantum information locally before distributing it across the network. This hybrid functionality simultaneously requires optical coherence, long-lived spin states, and reproducible integration with photonic and microwave control infrastructure. Scalability in this regime favors modular architectures, where identical nodes can be fabricated, characterized, and interconnected, rather than monolithically scaled systems that rely on uniform global control. With this, implementing large-scale computations will still place high demands on efficiency, fidelity, and rate. In addition, it will require very large numbers of qubits, which poses severe engineering challenges, including crosstalk mitigation and thermal management for cryogenic operation.

Because the distances between nodes in modular computing systems will be much shorter than those in a quantum repeater, photon emission times are more relevant in this application. The clock speed of current and envisioned quantum computing architectures may lie in the range of approximately $\SI{100}{\nano\second}$, corresponding to node separations up to $\SI{25}{\meter}$ of optical fiber. This timescale can be achieved with all emitters discussed in this review, at least with moderate, realistic increases in the Purcell factors (c.f. Tab.~\ref{tab:emittersincavity}). Again, increasing the emission rate below $\lesssim\SI{10}{\nano\second}$ will not enable higher clock rates owing to the signaling time; instead, it would only unnecessarily complicate controlling the optical qubit.

As an alternative, distributed quantum computers may also be implemented by optically linking quantum processors operating at microwave frequencies, such as quantum dots~\cite{vandersypen_interfacing_2017} or superconducting qubits. To this end, efficient microwave-to-optical converters are required, and different hardware platforms are being explored~\cite{lauk_perspectives_2020}. Among them, spin-bearing photon emitters have been proposed to simultaneously offer high efficiency and a large-enough bandwidth~\cite{williamson_magnetooptic_2014, khalifa_robust_2025}. To implement corresponding transducers, one requires microwave and optical resonators with high quality factors, as well as low inhomogeneous broadening of both optical and spin transitions. Currently, the measured values for erbium dopants~\cite{gritsch_narrow_2022, rinner_quantum_2026} or T centers~\cite{higginbottom_memory_2023} suggest that further improvements will be required to enable efficient transducers.

\subsubsection{Photonic quantum computing}

In addition to the hybrid spin-photon devices described above, quantum computers can also be realized in all-photonic platforms~\cite{rudolph_why_2017, slussarenko_photonic_2019}, which require only efficient sources and linear optical elements, ideally on-chip and with low loss. While current experiments often use nonlinear optical elements to generate the required photons, this approach incurs significant overhead. Instead, using single-photon emitters would dramatically reduce the number of required components, provided they achieve a very high degree of photon indistinguishability, placing stringent demands on single-photon purity, spectral stability, lifetime control, and emitter tunability. A recent study found that silicon waveguides may be too lossy to realize all-photonic processors based on heralded sources and probabilistic fusion~\cite{alexander_manufacturable_2025}. However, implementing deterministic single-photon generation and quantum gates using matter-mediated photon interactions~\cite{hacker_photonphoton_2016} can alleviate the stringent loss requirements. Thus, waveguide losses can be reduced, which may facilitate exceeding error correction thresholds~\cite{alexander_manufacturable_2025}.

In the context of all-optical quantum computing, maximizing the emission rate seems desirable when considering only clock speed. However, if the emission happens on too fast timescales, resonant optical excitation requires ultrashort pulses and thus becomes technically more challenging. Also, it demands higher powers, which may entail laser-induced dephasing, as observed with T centers~\cite{zhang_laser-induced_2025, bowness_laser-induced_2025} and Er dopants~\cite{frueh_spectral_2026}. Furthermore, two-photon emission is difficult to avoid with fast emitters~\cite{fischer_signatures_2017}, hampering fidelity. In summary, photonic quantum computers will likely aim for emission timescales of several nanoseconds, which are compatible with on-chip Er:Si quantum memories~\cite{rinner_quantum_2026}, potentially offering further scaling advantages: While in principle, photonic quantum computing architectures do not require spin-photon interfaces for memory and local processing, in practice, the resource overhead can be significantly reduced if they can be integrated into future devices. Again, corresponding approaches will require very high efficiencies and fidelities.

\subsection{Position of silicon-based platforms}

Silicon-based platforms occupy a unique position across the mentioned applications. Silicon offers unmatched maturity in nanofabrication, foundry-scale photonics, and low-loss microwave integration, enabling large-scale and reproducible device architectures. These strengths are particularly relevant for multiplexed quantum network nodes, scalable quantum processors, and chip-scale sensing systems. At the same time, the high purity of silicon crystals has enabled long spin coherence times, up to several hours for nuclear spins in electrically controlled dopants~\cite{saeedi_room-temperature_2013} and several seconds on the electronic spin transitions in isotopically purified bulk crystals~\cite{tyryshkin_electron_2012}. This has paved the way to high-fidelity spin control, with two-qubit gates above the threshold of common error-correction codes~\cite{noiri_fast_2022}. Initial experiments with T centers~\cite{photonic_inc_distributed_2024, song_entanglement_2025} and erbium dopants~\cite{gritsch_optical_2025, berkman_millisecond_2023} observed somewhat shorter coherence; nevertheless, this is not expected to pose a limitation to the targeted applications.

In contrast, silicon’s relatively narrow bandgap and complex defect landscape may impose constraints on the achievable optical coherence and spectral stability that must be addressed to meet application-driven requirements. In the following, we will examine how these constraints manifest at the emitter and materials level, and how different classes of silicon-based quantum emitters navigate the resulting trade-offs.

While multiplexed quantum sensing may not put hard requirements on the optical properties, the other mentioned applications require coherent photon emission, spectral reproducibility, and strong light--matter interaction to enable high efficiency and fidelity in remote entanglement. Quantitatively, one needs to achieve spin-photon interfaces with $C \gg 1$, which is still an open challenge in silicon. With T-centers, the limited reported cooperativities of $C\lesssim 0.1$~\cite{photonic_inc_distributed_2024, komza_multiplexed_2025} so far hamper the rate and fidelity of remote entanglement~\cite{photonic_inc_distributed_2024}. With Er:Si, $C \approx 1$ has been demonstrated~\cite{gritsch_optical_2025, frueh_spectral_2026}, mainly enabled by the narrower homogeneous linewidth~\cite{gritsch_narrow_2022, berkman_observing_2023}. Clearly, in both systems, further improvements of $C$ will be required for many applications.

Rather than being constrained by a single universal bottleneck, different silicon-based emitters may face distinct limitations, leading to complementary scaling pathways. For Er:Si, to further improve $C$, one will likely target stronger Purcell enhancement, as this would simultaneously reduce the lifetime from the current $\lesssim \SI{1}{\micro\second}$, which would limit the repetition rate in distributed quantum computing operations. To increase $F_\text{P}$, one will likely first try to enhance the resonator's $Q$. Recent measurements suggest that a tenfold improvement should be feasible without being limited by absorption losses~\cite{frueh_spectral_2026}. Unfortunately, increasing $Q$ reduces the operating bandwidth; thus, active resonator tuning will become increasingly relevant in Er:Si. Instead of increasing $Q$, a reduction of $V$ may also be considered. However, with the current emitter-integration procedure based on implantation, most devices with mode volumes below $\left( \lambda/n \right)^3$ would not contain a single dopant in the preferred sites. Post-selection would be possible, but inefficient. Thus, improving the integration yield will likely be a key target of future research.

\begin{figure*}[tb] 
\includegraphics[width=2.0\columnwidth]{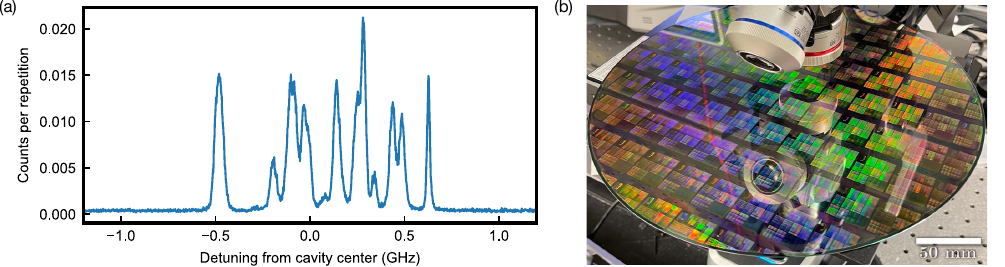}
\caption{     \label{fig:multiplexing}
\textbf{Spectral and spatial multiplexing.} (a) Spectral multiplexing. Pulsed resonant fluorescence spectroscopy is performed on a nanophotonic resonator with embedded erbium dopants. When scanning the excitation laser detuning, many distinct peaks are observed, each originating from a single emitter. The spectral diffusion linewidth of the dopants is much smaller than their emission frequency spread, enabling the spectrally selective excitation of individual emitters in the same device. Adapted from Gritsch et al.~\cite{gritsch_optical_2025}, Nature Communications, 16, 64, 2025; licensed under a Creative Commons Attribution (CC BY) license. (b) Spatial multiplexing is enabled by full-wafer photonic crystal fabrication in an
optimized 300 mm foundry process. A silicon wafer contains 64 complete reticles, each comprising millions of inverse-designed PhC cavities with free-space outcoupling. Adapted from Panuski et al.~\cite{panuski_full_arxiv_2022}, arXiv:2204.10302, 2022; licensed under a Creative Commons Attribution (CC BY) license.}
\end{figure*}

For color centers with their faster emission, one will likely aim to minimize the dephasing to achieve high values of $C$. Clearly, this would also benefit Er:Si. Recent measurements in both platforms suggest that dephasing is dominated by laser-induced charge noise~\cite{zhang_laser-induced_2025, bowness_laser-induced_2025, frueh_spectral_2026}. Thus, one may try to find an emitter with a symmetry that leads to a small sensitivity to electric and, possibly, magnetic fields, see Sec.~\ref{sec:introduction}. However, all color centers described in this review, and erbium dopants in all sites of known symmetry~\cite{vinh_photonic_2009, holzapfel_characterization_2025} belong to polar point groups, which entails a non-zero Stark shift. Reducing their dephasing will thus require stabilizing their charge environment. This may be achieved by optimizing materials and the nanofabrication procedure, as well as by surface passivation. Alternatively, or in addition, strong electric fields may be applied to the emitters, e.g., using integrated diode structures, as pioneered with quantum dot devices~\cite{lodahl_interfacing_2015} and color centers in other materials~\cite{anderson_electrical_2019}. The improvements possible in silicon devices still require experimental investigation.

\subsection{Strategies for upscaling}

So far, all experiments on single-photon emitters in silicon have been performed on isolated emitters in a single device. However, most of the applications mentioned in Sec.~\ref{subsec:Applications} require control over a large number of emitters --- the larger, the better. This is possible by spatial and spectral multiplexing, which will be discussed in the following.

\subsubsection{Spectral multiplexing} \label{sec:spectral_multiplexing}

In a spectral-multiplexing strategy, instead of separating emitters in space, multiple emitters are operated in the same device. To still control them individually, one can exploit the inhomogeneous distribution of their optical transition frequencies, which can exceed the SD linewidth of individual emitters. In this case, resonant laser excitation of one emitter may not significantly disturb the others, enabling parallel control with high fidelity. 

In principle, this approach can be applied to both color centers and erbium dopants. In other materials, hundreds of emitters have been spectrally resolved in the same resonator~\cite{ulanowski_spectral_2024}, and parallel spin readout and control \cite{chen_parallel_2020} as well as multiplexed entanglement generation~\cite{ruskuc_multiplexed_2025} have been demonstrated. 

In silicon, the inhomogeneous broadening of most emitters is below $\SI{10}{\giga\hertz}$, enabling optical connections by photon interference with temporally resolved detection. However, the spectral diffusion observed with individual color centers so far is of the same order of magnitude (see Tab.~\ref{tab:emittersincavity}), which may hamper a clear spectral separation of multiple emitters. In contrast, the narrower SD linewidth of erbium dopants allows resolving around ten emitters in a single nanophotonic resonator, all located within a subwavelength-scale mode volume, see Fig.~\ref{fig:multiplexing}(a).

In such nanophotonic resonators, the multiplexing capacity is limited by two aspects: First, their high $Q$ factor restricts the operational bandwidth unless a fast-tuning mechanism is implemented. Second, confining many emitters in a small volume requires a high density, which can spoil the SD linewidth and increase laser-induced spectral diffusion~\cite{frueh_spectral_2026}. Quantifying the limitations imposed by these considerations will need further experimental work.

\subsubsection{Spatial multiplexing} \label{sec:multiplexing}
The second strategy for the parallel operation of many emitters is spatial multiplexing. Here, one would simply place many identical devices next to one another on the same chip. In addition, photonic links between the devices can enable operating several of them in parallel, potentially in different and even in remote cryostats. This approach requires a high yield with minimal fluctuations across devices and thus clearly benefits from the maturity of silicon nanofabrication. Still, the resonance frequency of nanophotonic cavities exhibits typical frequency fluctuations in the range of $\lesssim \si{\tera\hertz}$. This can be overcome by several approaches. First, several resonators can be attached to a single feed waveguide~\cite{komza_multiplexed_2025}, increasing the likelihood that at least one operates at the desired frequency. Second, post-fabrication oxidation can be used to tune the resonators; this can be performed in a spatially selective way by laser-heating, even at the wafer-scale~\cite{panuski_full_2022}. Similarly, laser-induced heating can be used to selectively evaporate gas previously condensed on the resonators to tune their frequency~\cite{mosor_scanning_2005}. Other approaches for cavity tuning may also be enabled in future devices, either through mechanical actuation or by hybrid integration of electro-optical materials.

With this, thousands of devices may be operated on each photonic chip, as shown in ~\ref{fig:multiplexing}b. However, several engineering challenges arise when aiming at such large qubit numbers. For example, one needs to consider the finite cooling power of cryogenic systems, with typical values of several Watts at $\SI{4}{\kelvin}$ that drop to $\lesssim \SI{0.1}{\milli\watt}$ at $\SI{0.1}{\kelvin}$. For coherent resonant optical driving of a single emitter, one needs to avoid significant decay during excitation pulses that would lead to two-photon emission~\cite{fischer_signatures_2017}. This requires pulse durations about 3 orders of magnitude shorter than the lifetime. Owing to the quadratic scaling of the required power with Rabi frequency, corresponding pulses will contain $\sim 10^{6}$ photons, and thus an energy around $10^{-13}\,\si{\joule}$. Assuming that most of this light is absorbed, at a clock cycle of $\SI{10}{\mega\hertz}$, $10^6$ qubits can be repeatedly excited at $\SI{4}{\kelvin}$, but only 100 at $\SI{0.1}{\kelvin}$. Thus, one may prefer systems that enable higher operating temperatures.

The above estimate does not include the heat load from optical fiber connections, which will be negligible for on-chip routing and processing with only a small number of fiber interconnects. Systems that employ spin-photon interfaces will incur additional heat load from operations used to control the spin states. To minimize the required currents and thus to minimize heating, one may use superconducting transmission lines on the samples, as pioneered in other quantum systems \cite{blais_circuit_2021, kurizki_quantum_2015} and recently implemented with T centers~\cite{song_entanglement_2025}. In addition, resonant designs can be used to minimize the currents required on the leads. In this case, microwave heating may be dominated by dielectric losses in the SOI materials stack; this still needs to be quantified for samples containing photon emitters to estimate the resulting heating. In addition, a certain thermal load will result from the heat conduction of the wires; this can be minimized by optical pulse delivery, as studied previously in superconducting quantum systems~\cite{delaney_superconductingqubit_2022}.

Another engineering challenge in microwave control is cross-talk between adjacent lines. Mitigation strategies include magnetic field gradients~\cite{vandersypen_interfacing_2017} and combining microwave pulses with optical Stark shifts~\cite{chen_parallel_2020}. While all-optical spin qubit control~\cite{rogers_alloptical_2014} may be considered, in silicon, this may be impeded by the recently observed laser-induced spectral diffusion and spin mixing~\cite{zhang_laser-induced_2025, bowness_laser-induced_2025, gritsch_optical_2025, frueh_spectral_2026}. In addition to the heat load incurred in spin control, other challenges arise when aiming for very large qubit numbers. These will be discussed in the following sections.

\subsubsection{Deterministic emitter generation}

\begin{figure}[tb] 
\includegraphics[width=1.0\columnwidth,trim={0 0 0 0},clip=True]{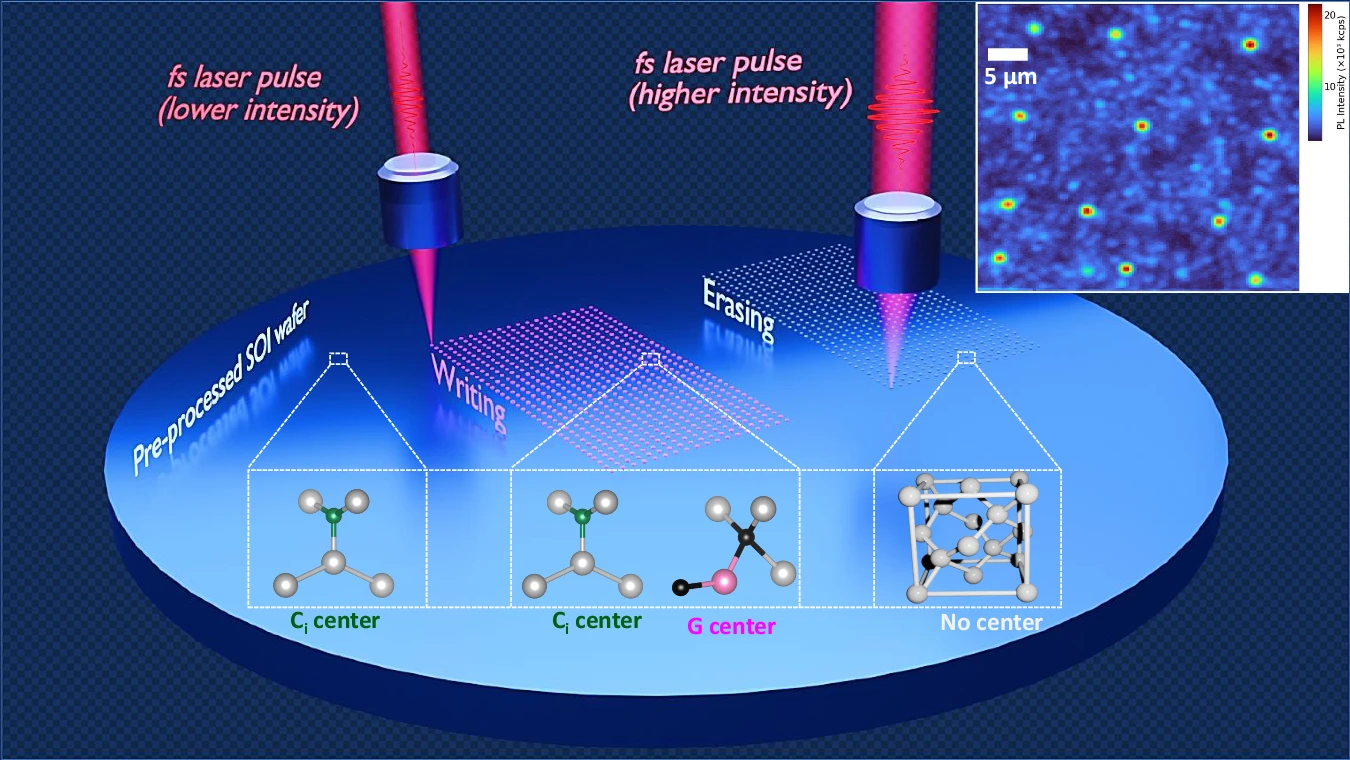}
\caption{     \label{fig:lasertrimming}
\textbf{Artistic representation of the femtosecond laser irradiation approach to locally write and erase color centers in SOI.} Insets: atomic structure of the laser-written color centers. Main figure: depending on its intensity, a femtosecond laser pulse locally forms (left) or erases (right) quantum emitters. Top right corner: a photoluminescence hyperspectral scan with laser pulse irradiation spots. Adapted from Jhuria et al.~\cite{jhuria_programmable_2024}, Nature Communications, 15, 4497, 2024; licensed under a Creative Commons Attribution (CC BY) license. }
\end{figure}

To implement systems with a large number of qubits, it is highly desirable to have a deterministic process in which each fabricated device contains a well-controlled number of emitters. However, previous approaches relied on ion implantation followed by global annealing. In this stochastic process, each device independently acquires a random number of emitters with a given probability. When spectral multiplexing is used, this may not be problematic. However, if the architecture requires exactly one emitter per device, the resulting Poissonian distribution would limit the yield to $\sim 37\,\%$. If a specific emitter orientation is required, this number would be reduced even further, potentially becoming prohibitively small.

A potential way to overcome this limitation is post-fabrication control over emitter formation, enabled by spatially selective laser annealing, as shown in Fig.~\ref{fig:lasertrimming}. Rather than relying on random activation, this approach enables emitters to be created, modified, or removed~\cite{prabhu_individually_2023, jhuria_programmable_2024} with high spatial selectivity. Importantly, the process does not create new atoms; instead, it locally reconfigures a pre-existing dopant or defect reservoir. 

Local laser annealing has been demonstrated to controllably create G center ensembles at predefined locations in silicon, paving the way for pre-patterning of large arrays of nanophotonic cavities followed by the incorporation of single or precisely counted emitters in each device \cite{andrini_activation_2024, jhuria_programmable_2024, gu_end--end_2025}. Beyond spatial control, laser-based approaches also offer a means to mitigate inhomogeneous broadening that arises from variations in local strain and defect environments~\cite{zhiyenbayev_scalable_2023, gu_end--end_2025}. For upscaling, one would set up a closed-loop workflow that combines laser annealing with automated optical and/or cryogenic characterization. Thus, arrays are first mapped to determine emitter occupancy, and the laser is then used to activate empty sites or remove excess emitters. 

Laser annealing may also find applications in systematic materials studies, particularly for emitter classes where the microscopic formation pathways remain poorly understood. As an example, a wide variety of lattice sites and charge configurations have been reported for erbium in silicon~\cite{kenyon_erbium_2005, vinh_photonic_2009, weiss_erbium_2021, gritsch_narrow_2022, berkman_observing_2023}, and the conditions required to reproducibly form optically coherent, potentially inversion-symmetric sites remain an active area of research. In this context, localized laser annealing provides a powerful experimental tool: by enabling site-resolved activation under controlled conditions, it allows large datasets to be acquired across a single chip, facilitating statistical studies of site formation, optical coherence, and spectral stability as a function of local processing parameters. Such high-throughput, spatially resolved exploration of annealing conditions is difficult with global thermal treatments and is currently lacking for erbium and many color centers in silicon.

The described deterministic strategies can be combined with modern foundry-scale nanophotonic platforms. Silicon photonic cavities with precisely engineered geometries and $Q/V>10^6$ can be fabricated at the wafer scale~\cite{panuski_full_2022}. By combining this large-scale photonic infrastructure with spatially resolved emitter activation and trimming, deterministic generation bridges the gap between materials-level control and system-level integration.

\subsubsection{Static frequency offsets and emitter tuning}

\begin{figure}[tb] 
\includegraphics[width=1.0\columnwidth]{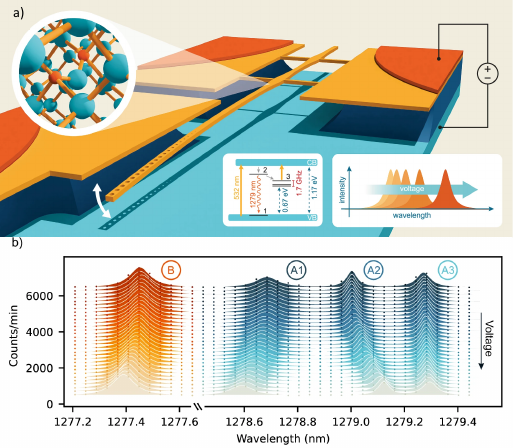}
\caption{\textbf{Strain tuning of a single silicon color center} (a) A ridge waveguide (orange bar) with a photonic crystal mirror on one end (holes) contains single G centers (left inset). It is freely suspended in a cryogenic vacuum. Application of an electric field to the device's electrodes (red) causes mechanical bending (white arrow), inducing strain at the emitter position. This leads to a tuning of the optical transition frequencies (bottom insets) with the applied voltage. (b) Experimentally observed shift of the optical emission frequencies of a single G center. Images adapted from Buzzi et al.~\cite{buzzi_spectral_2025},  	arXiv:2501.17290, 2025; licensed under a Creative Commons Attribution (CC BY) license. \label{fig:gtuning}
}
\end{figure}

In the spatial multiplexing described above, static frequency offsets between emitters are undesired. Also, other applications, in particular microwave-to-telecom transduction~\cite{williamson_magnetooptic_2014}, would require a strong reduction of the inhomogeneous linewidth. In nanophotonic silicon devices, the broadening is dominated by strain gradients. Those can originate from thermal expansion mismatches between the device layer and the surrounding materials, from crystalline defects generated during device fabrication and emitter formation, and from the different masses of the naturally occurring silicon isotopes, which lead to different zero-point fluctuation energies. With color centers, this produces ensemble linewidths of $\gtrsim \SI{10}{\giga\hertz}$, and $\gtrsim \SI{400}{\mega\hertz}$ for the shielded electrons in the 4f orbitals of erbium dopants. 

Thus, in all silicon emitters, the inhomogeneous linewidth exceeds the lifetime limit by many orders of magnitude, even in devices with the strongest Purcell enhancement. However, in bulk crystals with exceptional chemical and isotopical purity, linewidths of tens of megahertz have been achieved with color centers~\cite{chartrand_highly_2018, bergeron_silicon-integrated_2020}. Whether comparable values can be observed in nanostructured devices with their additional broadening mechanisms (outlined in Fig.~\ref{fig:overview}) remains a topic for future investigation. 

In case the inhomogeneous linewidth cannot be narrowed down, the detrimental effects of the random frequency distribution of the emitters can be counteracted by time-resolved detection~\cite{vittorini_entanglement_2014} or by frequency shifting either the light or the emitters. In practice, the last option may be most convenient. To this end, one may use controlled strain, as demonstrated with a single G center using microelectromechanical actuation~\cite{buzzi_spectral_2025}, see Fig.~\ref{fig:gtuning}. Alternatively, one can apply an electric field when the Stark coefficient is non-zero, which is the case for all emitters detailed in this review. Corresponding measurements have been performed on G center ensembles in diode structures~\cite{day_electrical_2024}, and T center ensembles in bulk crystals between electrodes~\cite{clear_optical-transition_2024}. A precise determination of the Stark coefficient would need the exact value of the applied electric field at the emitter position, which cannot be measured directly. Nevertheless, for the T center, a magnitude of the linear Stark coefficient between \qtyrange{3.5}{7.5}{\kilo\hertz \meter\per \volt} and quadratic Stark coefficients up to \qty{0.123}{\hertz \meter\squared \per \volt\squared} have been determined~\cite{clear_optical-transition_2024}. Theoretical calculations deviate from this result~\cite{alaerts_first-principles_2025}, which may hint at local field corrections from polarizable impurities.

With Er:Si, the Stark coefficient for two single dopants in unknown sites in electrically controlled devices has been estimated~\cite{zhang_single_2019} as \qty{0.1}{\kilo\hertz \meter\per \volt}. This is similar to observations in other materials in which erbium dopants are integrated in sites of polar symmetry~\cite{macfarlane_optical_2007}. Compared to T centers, the value is approximately two orders of magnitude smaller due to the mentioned shielding of the inner 4f electrons. The Stark coefficient of Er:Si in the reproducible sites A and B still has to be measured.

Given the observed Stark coefficients for both color centers and erbium dopants, the achievable tuning range with on-chip electrodes should be sufficient to tune two emitters to resonance, independent of their original frequencies within the inhomogeneous distribution. This paves the way for indistinguishable photon generation if the condition $C>1$ can be satisfied, enabling photonic quantum information processing.

\subsection{Scalable systems integration}

\begin{figure*} 
\includegraphics[width=2.0\columnwidth]{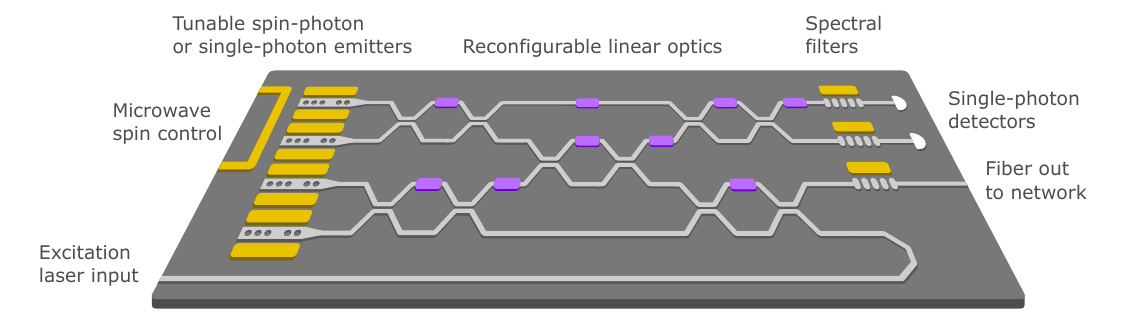}
\caption{     \label{fig:on_chip_optical_components}
\textbf{Schematic of a future quantum integrated photonic circuit based on silicon quantum emitters.} The schematic includes (left section) laser excitation, photon generation, and spin-photon interfaces; (middle section) linear optics for photon manipulation and photonic quantum-state generation; and (right section) spectral filtering and photon detection. Laser light from external sources is routed into the circuit to excite and control spin-photon interfaces and single-photon emitters. Photons from individual cavity-coupled single-photon emitters or microwave-controlled spin-photon interfaces with spectral tunability are coupled into low-loss photonic waveguides. Combining phase shifters and directional couplers enables arbitrary linear optical quantum operations. This is followed by spectral filtering and on-chip single-photon detection. One or more ports can be routed to other modules or to the fiber network for quantum communication. Multiple such modules can be fabricated on the same chip or connected via optical fibers.
}
\end{figure*}

Once deterministic arrays of emitters can be realized by multiplexing, deterministic integration, and frequency tuning, the prerequisites for a scalable quantum architecture may be fulfilled. Upscaling will then rely on efficient integration with photonic structures for both on-chip processing and off-chip connectivity. Previous experiments demonstrated the integration of color centers in diamond into hybrid photonic circuits \cite{wan_large-scale_2020, kim_hybrid_2020, starling_fully_2023, clark_nanoelectromechanical_2024, li_heterogeneous_2024}. Transferring these techniques to silicon photon emitters has the advantage that key functionalities can be realized in a single materials platform. In addition, their operating wavelengths in the telecom bands enable direct, low-loss interfacing with optical fibers and seamless compatibility with existing photonic infrastructure. This makes fibers not merely an output channel, but a core architectural element for scalable quantum systems.

At the chip level, nanophotonic resonators provide an interface between localized emitters and guided optical modes. Free-space optics remains a versatile complementary approach, particularly for parallel excitation and readout when emitters share the same frequency bin~\cite{li_heterogeneous_2024}. Alternatively, efficient off-chip couplers can be implemented without incurring prohibitive loss. Edge and grating couplers enable efficient transfer to standard single-mode fibers, while fiber arrays allow many optical channels to be addressed in parallel with high mechanical stability.

System-level demonstrations of packaged, multichannel cryogenic quantum memory modules~\cite{starling_fully_2023, zeng_cryogenic_2023}, illustrate that large numbers of emitters or qubits can be simultaneously controlled and read out within compact cryogenic environments when fiber-based interfaces are used. Routing photons off-chip also reduces reliance on cryogenic optical switching, which remains technologically challenging due to optical loss, power dissipation, and control complexity. 

Beyond static coupling, dynamic control of the emitter–photon interface can further enhance scalability. Microelectromechanical approaches have been used to modulate the coupling between spin qubits and photonic modes, enabling tunable interaction strengths and offering additional flexibility for optimizing entanglement rates and network performance~\cite{clark_nanoelectromechanical_2024}.

While the mentioned techniques can all be implemented in nanophotonic silicon circuits, hybrid integration strategies can extend these capabilities by combining silicon with other material platforms, such as silicon nitride, diamond, or two-dimensional materials. This allows the incorporation of additional functionality, including on-chip frequency conversion, filtering, nonlinear optical processing, active photon routing, and resonator tuning while preserving fiber compatibility, as shown in Fig.~\ref{fig:on_chip_optical_components}.

Overall, the integration of deterministic telecom emitters with silicon nanophotonics and fiber-based interconnects represents a central strength of the platform. By leveraging low-loss fibers for local, chip-scale, and long-range connectivity, telecom emitters in silicon naturally support modular, high-connectivity architectures that span atom–atom, chip–chip, and node–node links. This fiber-native approach enables scalable quantum communication, networking, and distributed quantum information processing while avoiding many of the scaling challenges associated with purely on-chip routing and cryogenic optical switching.

\section{Summary and Outlook}
Once the wonder material that enabled the information era, silicon is now emerging as a key material for the quantum information era. Complementary to all-electronic approaches~\cite{vandersypen_interfacing_2017}, recent works have explored photon emitters and spin-photon interfaces, whose connectivity opens new possibilities. In this review, we have summarized the state of the art in this young research field and identified Er dopants and T, G, W, and C centers as promising building blocks for future quantum technologies. We have described their optical and spin properties and summarized recent technological developments towards their integration into nanophotonic waveguides and cavities, localized formation, multiplexing, spectral tuning, and entanglement generation.

To enable upscaling to the large qubit numbers required for quantum applications, further improvements in the emitter properties are needed. We expect that future work will focus on i) investigating the remaining unknowns in the properties of the known emitters; ii) exploring whether new, yet-undiscovered emitters, offer improved performance; iii) achieving larger cooperativities by reducing the emitter dephasing and further improving the resonators; and iv) integrating emitters into functional nanophotonic circuits featuring excitation, routing, modulation, multiplexing, filtering, spin driving, and detection.

Even partial achievement of these goals will soon enable the demonstration of high-rate, high-fidelity optically mediated entanglement between remote spins. This will pave the way for fully functional quantum repeaters and, thus, global quantum networks~\cite{wehner_quantum_2018}. Further improvements of the system parameters will be required for quantum error correction, enabling scalable distributed quantum information processing between spins~\cite{simmons_scalable_2024}. In addition, novel approaches to all-photonic on-chip quantum information processing~\cite{alexander_manufacturable_2025, maring_versatile_2024} may come within reach in case high-cooperativity systems can be implemented. This includes the deterministic generation of Schrödinger cat states~\cite{hacker_deterministic_2019, le_jeannic_dynamical_2022}, photonic cluster states~\cite{lindner_proposal_2009, istrati_sequential_2020, thomas_efficient_2022, cogan_deterministic_2023}, and other many-photon entangled states~\cite{gonzalez-tudela_lightmatter_2024}. Combining these approaches with spin-photon interfaces enables the chip-based implementation of nondestructive photon detectors~\cite{reiserer_nondestructive_2013}, or deterministic photon-photon quantum gates~\cite{hacker_photonphoton_2016}. Finally, spin-photon interfaces with large qubit numbers will enable numerous applications in quantum simulation~\cite{noh_quantum_2016} and metrology~\cite{paulisch_quantum_2019}.

Thus, the maturity of silicon photonics and microelectronics, the exquisite quality of its isotopic purification, and the emission wavelength of embedded emitters position silicon as one of the most promising materials for the second quantum revolution. We expect that overcoming the remaining challenges will deliver on these promises and put silicon at the core of a wide variety of future quantum technologies.

\begin{acknowledgments}
The authors thank Christian Primavera for providing feedback on this manuscript.
C.E-H. acknowledges ERC DISQOVER (Grant No.\ 101219179) and the Dutch Research Council (NWO) through grants 601.QT.001 and NGF.1623.23.027.
A.G. acknowledges the support of the Quantum Information National Laboratory of Hungary, funded by the National Research, Development, and Innovation Office of Hungary (NKFIH) under Grant No.\ 2022-2.1.1-NL-2022-00004 and funding from the European Commission for the SPINUS (Grant No.\ 101135699) and QuSPARC (Grant No.\ 101186889) projects. 
A.R. acknowledges funding by the Deutsche Forschungsgemeinschaft (DFG, German Research Foundation) under the German Universities Excellence Initiative - EXC-2111 - 390814868 and via the individual grant agreement 559594594, and by the European Union (ERC project OpENSpinS, number 101170219). 
Views and opinions expressed are those of the authors only and do not necessarily reflect those of the European Union or the European Research Council. Neither the European Union nor the granting authority can be held responsible for them.
\end{acknowledgments}

\section*{Data Availability Statement}
Data sharing is not applicable to this article as no new data were created or analyzed in this study.

\end{document}